\documentclass[twocolumn,preprintnumbers,amsmath,amssymb]{revtex4}
\usepackage{epsfig}
\usepackage{amsmath}
\graphicspath{{figures/}}
\usepackage{epsfig}
\usepackage{amsmath}
\usepackage{color}
\usepackage{graphicx,epsfig}
\usepackage{color}
\usepackage[latin9]{inputenc}
\usepackage{float}
\usepackage{natbib}
\begin{document}

\title{The Madrid-2019 force field for electrolytes in water using TIP4P/2005 and scaled charges: extension to the 
ions F$^-$, Br$^-$,
I$^-$, Rb$^+$, Cs$^+$. }

\author{S. Blazquez$^2$}
\author{M. M. Conde$^1$}
\author{J. L. F. Abascal$^2$}
\author{C. Vega$^{2,a)}$}
\affiliation{$^1$Departamento de Ingenier\'{\i}a Qu\'{\i}mica Industrial y Medio Ambiente, Escuela T\'ecnica Superior de Ingenieros Industriales,
Universidad Polit\'ecnica de Madrid,
28006, Madrid, Spain}
\affiliation{$^2$Departamento Qu\'{\i}mica F\'{\i}sica I, Facultad de Ciencias Qu\'{\i}micas,
Universidad Complutense de Madrid, 28040 Madrid, Spain}

%\date{\today}
%
\begin{abstract}
In this work, an extension of the Madrid-2019 force field is presented.
We have added the cations Rb$^+$ and Cs$^+$  and the anions F$^-$, Br$^-$ and
I$^-$. These ions were the remaining alkaline and halogen ions not previously
considered in the Madrid-2019 force field. 
The force field, denoted as Madrid-2019-Extended,  does not include polarizability, 
and uses the  TIP4P/2005 model of water and scaled charges for the ions. 
A  charge  of $\pm$0.85 $e$ is assigned to monovalent ions.  The force field
	developed provides an accurate description of the aqueous solution
	densities over a wide range of concentrations up to the solubility
	limit of each salt studied. Good predictions of viscosity and diffusion
	coefficients are obtained for concentrations below 2 m.  Structural
	properties obtained with this force field are also in reasonable
	agreement with the experiment.  The number of contact ion pairs has
	been controlled to be low so as to avoid precipitation of the system at
	concentrations close to the experimental solubility limit.  A
	comprehensive comparison of the performance for aqueous solutions of
	alkaline halides of force fields of electrolytes using scaled and
	integer charges is now possible.  This comparison will help in the
	future to learn about the benefits and limitations of the use of scaled
	charges to describe electrolyte solutions.

\end{abstract}

\maketitle
a) Corresponding author: cvega@quim.ucm.es

\section{Introduction}
   Life started from the seas and for this reason, water is the main component
   of the cells. However, water is not alone in the oceans, since they contain
   a certain amount of several salts (NaCl being the most abundant component).
   Not surprisingly ions are also present in living organisms.  Arrhenius was
   the first to suggest that when salts dissolve in water, individual ions are
   solvated by water, and they are able to conduct the electric current, so
   that they become denoted as electrolytes. 
   
      The solution of electrolytes in water has been studied intensively over
      the past century, both from an experimental point of view, and from a
      theoretical point of view. As students we learn the thermodynamics of
      electrolytes and the Debye-H\"{u}ckel theory describing their
      activities\cite{debye-huckel-law}.  The birth of computer simulations in
      the 1950's  brought a new route to study these type of systems. In the
      70's the first simulations of ionic systems were carried
      out\cite{doi:10.1080/00018737600101392,singer77,Heizinger1,heizinger2,Heizinger3}.
      Aqueous solutions of electrolytes require both a force field for the
      molecule of water and another one for the ions in water. In the 80's
      several force fields for water were proposed (TIP3P\cite{jorgensen83},
      TIP4P\cite{jorgensen83}, SPC/E\cite{spce}) that are still use widely
      nowadays. In those force fields the molecule of water is described by a
      rigid non-polarizable model, typically using a LJ center is located on
      the oxygen and  charges are located in the protons and in other places in
      the molecule.  In early 2000 new models of water were proposed, including
      TIP4P-Ew\cite{tip4p-ew}, TIP5P\cite{mahoney01} and
      TIP4P/2005\cite{abascal05b}.  A common feature of these three models is
      that they were able to reproduce the maximum in density of pure water. It
      has been shown that among the rigid non-polarizable models of water
      TIP4P/2005 is a quite decent one\cite{vega11,vega09}. When describing the
      force field for the ions, they are often described by a LJ center and a
      certain charge. The obvious option is to assign a charge of $\pm$Z e to the
      ions.  

      Computer simulations of electrolytes are not a fully mature area. In fact
      it is surprising that most of the simulation studies dealing with these
      substances do not consider the possibility of evaluating properties such
      as freezing depression, activity coefficients, salt solubilities, etc.
      However there has been significant progress over the last ten years and
      we anticipate that these properties will be computed and will bring some
      surprises. 

      There are a large number of models for different alkaline and
      alkaline-earth salts
      \cite{smith18,opls,str:jcp88,aqvist,dang:jcp92,beg:jcp94,SmithDang,roux:bj96,peng:jcpca97,wee:jcp03,Jorgensen,lam:jcpb06,
      ale:pre07,len:jcp07,joung08,gallo_anomalies,cal:jpca10,yu:jctc10,reif:jcp11,gee:jctc11,
      deu:jcp12,
      mao:jcp12,mam:jcp13,mou:jctc13,kiss:jcp14,kol:jcp16,elf:epjst16,pethes17,JCTC_16_2460_2020}
      but standard force fields do not have problems in general (when properly
      designed) in reproducing the experimental densities for a particular
      salt, as for instance NaCl. Some force fields such as that proposed by
      Smith and Dang\cite{SmithDang} were designed for just one electrolyte
      (NaCl in this case).  It is more difficult to find a force field able to
      describe simultaneously the properties of a large set of salts (i.e NaCl,
      KCl, MgCl$_2$ ...).  Among the fields of this type, the most popular is
      that proposed by Joung and Cheatham\cite{joung08} for three particular
      water models (TIP3P\cite{jorgensen83}, TIP4P-Ew\cite{tip4p-ew}, and
      SPC/E\cite{spce}).  Since TIP4P/2005\cite{abascal05b} is now regarded as
      a good potential model for water\cite{vega11} it seems reasonable to
      develop a force field of electrolytes based on this potential model of
      water. Dopke \textit{et al.}\cite{dopke} have recently shown that the
      force field proposed by Joung and Cheatham could also be used for
      TIP4P/2005 (by using Lorentz-Berthelot combining rules) provided that one
      uses either the parameters designed for SPC/E or TIP4P-Ew. We reached a
      similar conclusion in 2016 when considering NaCl\cite{ben:jcp16}.  In any
      case, although results were reasonable it seems logical to design a force
      field of electrolytes specifically for the TIP4P/2005 model of water. We
      started down this route in 2017 for NaCl\cite{ben:jcp17}, and continued
      in 2019 presenting a force field\cite{JCP_2019_151_134504} of some ions
      (Li$^+$, Na$^+$, K$^+$, Cl$^-$,Mg$^{2+}$, Ca$^{2+}$, SO$_{4}^{2-}$) using
      TIP4P/2005 water. The choice of the ions in our 2019 paper was not
      random, but we selected the most abundant ions presented in seawater
      (which are also the most common ions found in the cells). We have shown
      that in fact an extremely accurate description of the properties of
      seawater is possible by using this force
      field\cite{doi:10.1021/acs.jctc.1c00072}. The name of the force field
      designed for electrolytes in TIP4P/2005 was denoted as Madrid-2019. The
      purpose of this work is to extend the force field to the ions considered
      by Joung and Cheatham but not considered when proposing the Madrid-2019
      force field, namely the anions F$^{-}$, Br$^{-}$ and I$^{-}$ and the
      cations Rb$^{+}$ and Cs$^{+}$. The motivation for this is twofold. On the
      one hand to extend the applicability of the Madrid-2019 force field, and
      on the other to allow for a direct comparison between the force field
      proposed by Joung and Cheatham and the Madrid-2019 force field.  The
      difference between the Joung-Cheatham and the Madrid-2019 force field is
      not only the choice of the water model but there is also a conceptual
      idea that makes this comparison specially useful as we will explain
      later. 

     Besides the density there are some other properties that are of interest
     when modelling electrolytes.  For instance transport properties such as
     viscosities or individual diffusion coefficients of the ions and of water.
     Also the solubility could be of interest (as the solubility of a certain
     force field should not necessarily correspond to the experimental value).
     The overwhelming majority of the force fields proposed for electrolytes up
     to 2011 used integer values (in electron units) of the charge of the ions
     in solution.  How is their performance when describing the properties of
     these solutions? Let us briefly summarize the situation: 

   \begin{itemize} \item{ Regarding transport properties it was shown by Kim et
			   al.\cite{kim12} that all force fields underestimated
			   significantly the diffusion coefficient of water at
			   high concentrations. In other words, the force
			   fields overestimated the impact of the salt in
		   reduction of the diffusion coefficient of water. } \item{
				   Yue and
				   Panagiotopoulos\cite{doi:10.1080/00268976.2019.1645901}
				   and also ourselves\cite{JCP_2019_151_134504}
				   have shown that the viscosity of the
				   solution is dramatically overestimated at
				   high concentrations, sometimes having a
				   viscosity up to 2-4 times larger than in
				   experiments. Bearing in mind the
				   Stokes-Einstein
				   relation\cite{dill2003molecular}, this could
				   be expected given the impact of the salt on
			   the diffusion coefficient of water.} \item{ The
			   solubility of most of the force fields using integer
			   charges is quite low when compared to
			   experiment\cite{doi:10.1063/5.0012102}. In the
			   particular case of NaCl the majority of models give
			   a solubility too low by a factor of
			   2-10\cite{doi:10.1063/1.4926840,doi:10.1063/1.4906320,doi:10.1063/1.4959789,doi:10.1063/1.4964725,jiang15}.
			   However it is clear after the work of Tanaka and
			   coworkers\cite{JCTC_16_2460_2020} that when properly
			   designed the use of integer charges still allows for
			   a good description of the solubility. Calculating
			   solubilities via computer simulations is difficult,
			   and only feasible in the last 10 years. For this
			   reason all force fields proposed before 2011 did not
			   consider solubility as a property of interest.
			   Therefore although the use of integer charges should
			   not necessarily provoke a low solubility, it turns
			   out that this was the case for practically all force
			   fields previously designed using integer charges.}
		   \item { Related to the previous point is the finding that
			   for a number of force fields of electrolytes the
			   number of contact ion pairs (CIP) (i.e a cation in
			   contact with an anion in
			   solution\cite{doi:10.1021/acs.chemrev.5b00742,doi:10.1021/jp809782z,doi:10.1021/acs.jctc.7b00846,doi:10.1063/1.2186641})
			   was quite high and aggregation of ions (which can be
			   regarded as the initial step of precipitation) was
			   observed in many simulations even at concentrations
			   well below the experimental solubility\cite{Spohr}
			   (reflecting that the solubility of the force field
			   was well below the experimental one). Actually, ion
			   clustering has been reported for different salts
			   below its experimental solubility limit. This fact
			   can be seen in monovalent salts as
			   NaCl\cite{mou:jcp13,mou:jpcb12,ale:jcp09,ale:pre07},
			   KCl\cite{auf:jctc07}, divalent salts
			   CaCl$_2$\cite{mar:jcp18}, or even sulphates as
			   Na$_2$SO$_4$\cite{wer:jctc10} and
			   Li$_2$SO$_4$\cite{plu:jpca13}.} \end{itemize}

   How to go to the next generation of force fields for electrolytes? Ab-initio
   calculations can be useful as shown by Ding et \textit{al.}\cite{Ding3310},
   but these types of calculations are quite expensive from a computational
   point of view. Introducing polarizability is another possibility (see the
   work of Kiss and Baranyai\cite{kissB3k,kiss:jcp14}) and certainly further
   research will continue to appear in this area. Notice that including
   polarization does not guarantee good solubilities, as was shown for the
   Baranyai force field\cite{jiang15}. However it is possible to modify the
   parameters to improve their predictions\cite{kol:jcp16}.  First principles
   calculations and polarizable models will continue being developed in the
   future.  However there is a simple and cheap approach that could improve the
   performance without extra computational cost. The recipe is simple, and it
   amounts to using scaled charges for the ions (i.e force fields in which the
   charge of the ions is $\lambda$$e$ where $\lambda$ is a number smaller than
   one).  The use of scaled charges is common in simulations of ionic
   liquids\cite{C9CP04947A,doi:10.1021/je400858t,FILETI2014205,C0CP02778B,C2CP23329K}
   so exploring this route for electrolytes seems of interest. 

    Let us briefly describe the ``history'' of scaled charge models. This was
    first suggested by Leontyev and Stuchebrukhov\cite{leontyev09,leontyev10a,
    leontyev10b,leontyev11,leontyev12,leontyev14} , by realizing that the
    dielectric constant at high frequencies of non-polarizable models is 1,
    whereas for water its value is 1.78. With that in mind Leontyev and
    Stuchebrukhov suggested using a scaled charge of $q_{scaled} ($e$) =
    1/\sqrt{\epsilon_{\infty}} $ for the ions. This lead to a charge 
  of 0.75 Z $e$ for the ions in water. This suggestion was further expanded by Jungwirth and 
  coworkers\cite{plu:jpca13,koh:jcpb14,koh:jpcb15,dub:jcpb17,mar:jcp18}, and 
  they have developed a potential model for several ions using this value for the charge of the ions. 
  Kann and Skinner follow a somewhat different approach\cite{kan:jcp14}. If you want to recover the Debye-H\"{u}ckel law, 
  you should have a force field, describing accurately the density of water, and where the strength 
  of the Coulombic energy between ions at infinite dilution and infinitely large distances should be 
  identical in your model and in experiment. Many force fields of water do not reproduce the experimental 
  value of the dielectric constant of pure water, and that scaled charges should be used to recover the 
  Debye-H\"{u}ckel law. In the particular case of TIP4P/2005 this results in a value of 0.85 
  Z $e$ \cite{kan:jcp14}. 
  One of us has also pointed out that different charges may be needed to describe the potential energy 
  surface and the dipole moment surface\cite{vegamp15}. 
  Notice that for water models without charges, such as the mW\cite{doi:10.1021/jp805227c}, it is 
  possible to design a force field for electrolytes without charges, as is the route followed by 
  Molinero and coworkers and still obtain good results\cite{doi:10.1063/1.3170982}. 
Barbosa and coworkers\cite{fue:jpc16} have also used scaled charges. Other authors such as Wang\cite{li:jcp15} or van der Wegt\cite{doi:10.1063/1.5017101}
  have also explored the use of scaled charges.
  The effect of charge transfer has been also studied by Rick\cite{doi:10.1063/1.4736851,doi:10.1063/1.3589419,doi:10.1063/1.4874256,SONIAT201631}
  suggesting
  that part of the charge of the ions is transferred to the adjoining water molecules,
  as also shown later by Yao et \textit{al.}\cite{Berkowitz}.
This charge transfer implies that the charge of the ions is not unity and has also been confirmed by quantum calculations\cite{doi:10.1021/acs.jctc.7b00846}.
  The community using scaled charges is growing and
  Jungwirth and coworkers have summarized the situation in a couple of review papers\cite{doi:10.1063/5.0017775,doi:10.1021/acs.jpclett.9b02652}.

    In our previous studies in 2017\cite{ben:jcp17} and in 2019\cite{JCP_2019_151_134504} we have shown that the use of scaled charges, improves 
  the description of the solubility (at least for NaCl), and the same is true for the viscosity 
  and the diffusion coefficient of water. Yue and Panagiotopoulos have presented further evidence of that\cite{doi:10.1080/00268976.2019.1645901}. 
  In fact, we have shown recently that more complex phenomena such as the salting out effect of methane can
  be quantitatively described with the use of scaled charges\cite{FPE_2020_513_112548}.
  Thus scaled charges seem to improve the description of aqueous solutions of electrolytes. 
  Scaled charges can not describe everything. In fact they can not describe neither the melt, the solid, 
  the vapor-liquid equilibrium of the molten salts, or the kinetics of precipitation\cite{doi:10.1063/5.0012065,doi:10.1063/5.0012102}. 
  Scaled charges are useful only when describing the properties of aqueous solutions of electrolytes 
  but not in problems where the system does not contain water (or has a very small amount). 

    To sum up in this paper we extend the Madrid-2019\cite{JCP_2019_151_134504} force field to other monovalent cations and anions belonging 
  to the alkaline group and to the halogens. By doing that we now have a set of parameters for the same set 
  of ions selected by Joung and Cheatham\cite{joung08}, and ready for their use with the TIP4P/2005\cite{abascal05b} model of water that 
  was not considered in the original parameterization of Joung and Cheatham. A consequence, is that now 
  a face to face  comparison of a good force field designed using integer charges, and a force field designed
  using scaled charges is possible. We hope that this comparison will shed light on the possible 
  advantages and disadvantages of the use of scaled charges for modelling electrolyte solutions.

\section{ Extending the Madrid-2019 force field to other electrolytes }

 In this work, an extension (to other ions) of the recently developed
Madrid-2019\cite{JCP_2019_151_134504} force field
of scaled
charges has been carried out. 
Madrid-2019\cite{JCP_2019_151_134504} 
force field was initially proposed to describe the most common salts present in seawater. 
In this work we extend the model to the rest of the halogens
that remained to be studied (F$^-$, Br$^-$ and I$^-$) and to
the rest of the cations of the alkaline group (Rb$^+$ and Cs$^+$).
When designing force fields, certain properties are used as the target, and the values of the parameters 
of the potential are obtained so that these properties are reproduced. 
 
  Now we describe in detail the philosophy of the Madrid-2019 force field, and the approach we used
  to obtain the parameters of the potential. 
  Obtaining parameters of a force field is a difficult job. It requires patience, a good design, and some trial and 
   error. In the future machine learning techniques could certainly help in obtaining the best set of parameters 
   for a certain potential model. It should be recognized from the very beginning that the only way to reproduce
   experimental results for all properties would be to solve the Schr\"odinger equation exactly, and to include nuclear
   quantum effects. If this is not done, your approach will not be able to reproduce ``everything'' and certain properties
   will be reproduced but other properties not. 
 These are the main characteristics of the Madrid-2019 force field and the description of how the parameters were obtained:  
 \begin{itemize}
	 \item{ Water is described by the TIP4P/2005 model. Ions are described by a Lenard-Jones(LJ) center and a certain charge. 
        In the case of the Madrid-2019 force field, the charge (in electron units) is scaled and its value is 0.85
		 for monovalents ions and 1.70 for divalent ions. The force field requires one to know the LJ parameters 
		 ($\sigma$ and $\epsilon$) of the interaction of a certain ion with the rest of species of the system. }

 \item { The Lennard-Jones interaction between an ion and water is described by the ion-oxygen interaction (i.e there 
	 is no ion-hydrogen interaction). The reason for that is that in the TIP4P/2005 model of water there is no LJ center 
		 on the hydrogen atoms of the molecule of water. }

 \item{ Lorentz-Berthelot (LB) rules are in general not used. Thus, the parameters of the LJ  ion-oxygen interaction 
	  and between a cation and an anion are not obtained via the LB combination rule. However, the interaction 
		 between two different types of cations and/or two different type of anions  
		 (required to study systems having simultaneously several salts as in the case of seawater) are usually given by LB combining rules. }

 \item { The force field is transferable. The parameters of the interaction of an ion with other species (water, other ions ...) 
	  are always the same regardless of the chemical composition of the system. }

 \item{ In this work, the parameters for the interaction between the ions Cl$^{-}$, Li$^{+}$, Na$^{+}$, K$^{+}$, Mg$^{2+}$, Ca$^{2+}$ with water and the crossed 
	  interactions between these ions is taken from our previous work in which the force field Madrid-2019\cite{JCP_2019_151_134504} was 
		 introduced. This is an advantage as it reduces the number of parameters to be determined. }

 \item{ The density of solutions is always considered as a target property. The density of ionic solutions is known with 
	  high accuracy from experiments, and we always use it as the target property (at room temperature and pressure). }

 \item { The LJ parameters of the ion-water interaction were determined by using the experimental densities for low 
	  concentrations (typically up to  1-2 molal , i.e less than 1-2 mols of salt per kilogram of water). The reason for that 
	  is that at low concentrations the properties of the solution are given mainly by the ion-water interaction
          and the impact of ion-ion interaction at low concentrations is quite small. As an example to determine the 
          properties of bromide salts, we shall consider the experimental densities of LiBr, NaBr, KBr, MgBr$_2$, CaBr$_2$ and will determine 
		 the parameters of the Br$^{-}$-water interaction using the experimental densities (up to 1-2 $m$) as target. Notice that for these
          salts the cation-water interaction is already available for the Madrid-2019 force field, so we only 
          need to determine the LJ parameters of the Br$^{-}$-water interaction.}

	 \item{ Obtaining the Br$^{-}$-Br$^{-}$ and  Br$^{-}$-cation interactions (we shall use Br as an example to illustrate how the force field 
	  was obtained) is a somewhat more involved process. Again we shall consider the experimental properties of LiBr, 
	  NaBr, KBr, MgBr$_2$, CaBr$_2$  and used the Madrid-2019 for the cation-cation 
          interaction. Thus one needs to determine the Br$^{-}$-cation interactions and the Br$^{-}$-Br$^{-}$
		 interaction. 
          This was done using three target properties. The density of the melt should be reproduced (i.e we used as the target 
	  a density 20-25 per cent lower than the experimental one). The experimental 
	  density of the ionic solutions at high concentrations (close to the experimental solubility limit) should be reproduced.
		 The third constraint is the number of contact ion pairs (CIP).
In the study of aqueous electrolyte solutions it is
important to evaluate the number of CIP.
High  values of CIP  indicate (indirectly)  cluster formation
and/or precipitation of the salt. Benavides et \textit{al.}\cite{ben:mp17} suggested
in a previous work that for 1:1 electrolytes with solubility lower
than 11\,m  the number of CIP
must be below 0.5 to be sure that precipitation and/or aggregation of ions has
not occurred. This rule will be important in the majority of the salts of this work.
(we will discuss later the number of CIP for exceptional cases such as salts with extremely high solubilities).
		 Thus, obtaining Br$^{-}$-Br$^{-}$
		 and Br$^{-}$-cation interaction 
	 is the more difficult step.
The reader may wonder why we did not try to reproduce the experimental densities
		 of the melt with scaled charges. 
We explored this approach but found that often the salt in solution had a large number
		 of CIP and in some cases precipitated.  For instance for NaBr we were able to find a set
		 of parameters describing the melt quite well (using scaled charges),
		 but with these parameters the number of CIP in the aqueous solution was large (3.4 at 8 m) and the salt precipitated spontaneously. 
Thus it seems that to obtain a correct balance between ion-water and ion-ion interactions
when using scaled charges, the target density of the melt should be somewhat smaller (20-25 $\%$)
than the experimental value.}

 \end{itemize}
 
	  The last step described above, i.e obtaining for instance Br$^{-}$-Br$^{-}$ and Br$^{-}$-cation interactions requires some further 
	  clarifications.  The density of the melt is used as a target property. We prefer the melt with respect to the ionic solid 
	  as when using empirical force fields the mechanical stability of the experimental solid structure is not 
	  guaranteed. Besides this, in the melt one is sampling anion-anion, cation-cation and anion-cation interactions. 
         However the target density was not the experimental density of the melt, but a density typically around 20-25 per cent lower than
         the experimental one.  The reason for that is that we recognize from the very beginning that scaled charges are not
         adequate to reproduce the properties of pure ionic systems as has been pointed out by Panagiotopoulos and coworkers\cite{doi:10.1063/5.0012065,doi:10.1063/5.0012102}. 
	 We observed that when using scaled charges for the melt one obtains a density 20-25 per cent lower than 
	 the experimental value and a reasonably low number of CIP
	 avoiding precipitation of the system. However, when replacing the charge with integer values (while keeping the LJ parameters)
	 the experimental density of the melt is reproduced. 
         Our approach will allow one in the future to design and develop a force field in which the charge of the ions
	 is sensitive to the local environment, within the spirit of polarizable models. 

         Also it is interesting to point out that ideally 
	 the solubility of the salt should be considered as a target property. If the solubility is low,
         then the ions will tend to cluster in solution (or even precipitate spontaneously) and the results of the 
         force field would be unreliable. 
	 The calculations of the solubility of a salt is painful from a computational point of view
	 and evaluating it for several trial values of the potential parameters would be beyond current computational 
	 limits. However there is a rather simple approach, at least to avoid low values of the solubility.

         We have shown in the past for salts with solubilities smaller than 11 m, the number of CIP at the solubility limit 
	 of the model is less than 0.5\cite{ben:mp17}. Thus having a number of CIP larger than 0.5 is a warning which indicates that we are not that far from 
	 the solubility of the model and the risk of spontaneous precipitation exists 
         (although since nucleation is an activated process, precipitation may occur at concentrations
	 several times higher than the solubility limit, as it happen for the JC-SPC/E model of NaCl\cite{https://doi.org/10.1002/jcc.25554,doi:10.1063/1.5084248}).
	 Therefore we will always force the 
	 force field to have a number of CIP less than 0.5 at the experimental value of the solubility (whenever below
		 11 m). For salts with higher solubilities we will discuss later the number of CIP. 

        In summary the main goal of the Madrid-2019 force field is to reproduce thermodynamic and transport properties
	of ionic solutions sacrificing somehow the properties of the melt and/or the solid. The main property of the solution that can not 
	be reproduced by the introduction of scaled charges is the free energy of hydration, which will be low compared 
	to experiments (although it can be corrected in a theoretical way\cite{FPE_2020_513_112548}).
	Nikitin and Frate\cite{https://doi.org/10.1002/jcc.26021} pointed out
	that the calculation of the total free energy of hydration, $\Delta$G is better (when including a theoretical correction)
	by using scaled charges. They consider
	that the total hydration energy can be divided in two terms. One is the calculated in simulations  $\Delta$G$_{MD}$ 
	and the other one is a theoretical correction denoted as the electronic polarization ($\Delta$G$_{el}$).
	That resembles the situation  for water 
	of models like SPC/E\cite{spce}, TIP4P-Ew\cite{tip4p-ew} or TIP4P/2005\cite{abascal05b} that sacrifice the enthalpy of vaporization of water as a target 
	property (in contrast with TIP3P\cite{jorgensen83}, TIP4P\cite{jorgensen83} and TIP5P\cite{mahoney01}) to obtain an overall better description of its properties. 
	Notice though that activity coefficients, osmotic pressures and a number of properties of solution can be 
	still be reproduced even though the absolute value of the hydration free energy is not reproduced. 
	This is so because what really matters is the variation of the Gibbs free energy of the system with the 
	addition of salt, rather than the absolute values of the Gibbs free energies. Transfer of a salt from vacuum to water,
	will not be described properly by using scaled charges 
	but this is not a big problem as ions are usually not found either in vacuum (due to 
	their low solubilities) in problems of practical interest. Notice, that
	Vazdar et al.\cite{doi:10.1021/jz300805b} showed in 2012 that using scaled charges
	(i.e. an electronic continuum correction) allow to describe reasonably well the 
	 hydrophobic oil/water interface. In general, they proposed that interfaces with 
	 no electronic discontinuity can be reasonably described by using scaled charges.

   We shall now present the parameters of the Madrid-2019 extended to the new ions 
   (F$^-$, Br$^-$, I$^-$, Rb$^+$, Cs$^+$). We shall denote this as Madrid-2019-Extended. 
The interaction between atoms can be described by two different contributions:
The first one, an electrostatic (coulombic) contribution and the second
one a van der Waals interaction represented by the LJ potential:
 
 \begin{equation}
u(r_{ij})= \frac{1}{4\pi \varepsilon_{0}}\frac{q_{i}q_{j}}{r_{ij}} + 4\epsilon_{ij}\left[\left(\frac{\sigma_{ij}}{r_{ij}}\right)^{12}
- \left(\frac{\sigma_{ij}}{r_{ij}}\right)^{6} \right],
\end{equation}
where, $q_{i}$ is the ionic charge, $\varepsilon_{0}$ is the vacuum permittivity, $\epsilon_{ij}$ 
the well depth energy of the LJ potential, and $\sigma_{ij}$ the LJ diameter.

 Let us now describe   the TIP4P/2005 model of water 
developed by Abascal and Vega\cite{abascal05b}. In this model, based on the TIP4P water proposed 
 by Jorgensen et al. \cite{jorgensen83},  water molecules have four atoms, two hydrogens with charge q$_{_{H}}$,
 one oxygen which is a LJ site but has no charge and one point M, near the oxygen atom on the symmetric axis, without mass but with charge q$_{_{M}}$.
 The geometry of water molecules for the TIP4P/2005 model can be described
 by the following parameters: oxygen-hydrogen distance, d$_{OH}$=0.9572 \r{A}, oxygen-M distance, d$_{OM}$=0.1546 \r{A} and angle H-O-H=104.52$^{\circ}$. TIP4P/2005 parameters are collected in Table \ref{tabla-agua}.
 
 \begin{table}[H]
\caption{Force field parameters for water TIP4P/2005 parameters  
from Ref\cite{abascal05b}.}
\label{tabla-agua}
  \begin{center}
    \begin{tabular}{ c c c c c c c c c c c }
\hline
\hline
Molecule & & Charge & & $\sigma_{ii}$ & & $\epsilon_{ii}$ & & \\
    & & (e) & & (\r{A}) & & (kJ/mol) & & \\
\hline
TIP4P/2005\\
O  & & 0  & & 3.1589  & &   0.7749 \\
H & & 0.5564  & &  & &  \\
M & & -1.1128  & &  & &  \\

\hline
\hline
    \end{tabular}
  \end{center}
\end{table}

We proceed now to present the Madrid-2019-Extended model parameters, which are collected in Tables \ref{tab_sig} and \ref{tab_eps}. 
We show only the parameters obtained in this work, (when for a certain interaction one
reads LB, it means that the interaction has been obtained from
the application of the LB combining rule).
The salts developed in Madrid-2019 original model\cite{JCP_2019_151_134504}
are denoted as Madrid-2019.

 \begin{table*}[!hbt] \centering
\caption{Lennard-Jones parameters $\sigma_{ij}$ (in \AA)
for the ions Li$^{+}$, Na$^{+}$, K$^{+}$, Rb$^{+}$, Cs$^{+}$,
	 Mg$^{2+}$, Ca$^{2+}$, F$^{-}$, Cl$^{-}$, Br$^{-}$, I$^{-}$ and SO$_{4}^{2-}$. O$_{w}$ and O$_{s}$ are the water an sulfate oxygens.}
\label{tab_sig}
  \begin{center}
 \resizebox{18cm}{!}{
    \begin{tabular}{c c c c c c c c c c c c c c c}
\hline
\hline
	    & F$^{-}$ & Cl$^{-}$ & Br$^{-}$ & I$^{-}$  & Li$^{+}$ & Na$^{+}$ & K$^{+}$ & Rb$^{+}$ & Cs$^{+}$     & Mg$^{2+}$     & Ca$^{2+}$      &O$_{w}$ & S & 
		    O$_{s}$ \\
\hline
	    F$^{-}$ & 3.78982 &   LB &    LB &    LB &    2.84540 &        LB &    3.46250 &        3.57250 &        3.94550 &        LB &    LB &    3.77450 &  LB  & LB \\

Cl$^{-}$     & &      Madrid-2019     & LB &  LB &    Madrid-2019 &   Madrid-2019 &   Madrid-2019 &   3.99642 & 4.31854 &     Madrid-2019     & Madrid-2019 & Madrid-2019& Madrid-2019  & Madrid-2019 \\

	    Br$^{-}$     &        &        &  4.82525     & LB & 2.61450 &        3.38500 & 3.79879 &       3.91725 & 4.33408       & 2.65519       & 3.67052       & 4.19850 & LB  & LB  \\

I$^{-}$     &        &        &        &     5.04975 &        3.20470  & 3.64658 &     4.00550 &        4.10288 &       4.43790  & 2.82707 &     3.94181 &       4.34950  & LB & LB  \\

Li$^{+}$     &  &    &    &     & Madrid-2019 &       Madrid-2019     & Madrid-2019 & LB &    LB &    Madrid-2019 &   Madrid-2019 &   Madrid-2019 & Madrid-2019  & Madrid-2019 \\

Na$^{+}$     &        &  &   &   &  & Madrid-2019     & Madrid-2019 & LB &    LB &    Madrid-2019     & Madrid-2019 & Madrid-2019 & Madrid-2019  & Madrid-2019  \\

K$^{+}$      &        &        & &     &    &  & Madrid-2019 &        LB &    LB      & Madrid-2019 & Madrid-2019 &   Madrid-2019 & Madrid-2019  & Madrid-2019 \\

Rb$^{+}$      &        &        &  &     &    &   &   & 2.99498 &     LB &    LB &    LB &    3.54350 & LB  & 3.40000 \\

Cs$^{+}$      &        &        &    &     &     &     &    &    & 3.521013 & LB      & LB &  3.66290 & LB  & LB \\

Mg$^{2+}$     &        &        &        &   &   &    &    &   &  & Madrid-2019 &      Madrid-2019     & Madrid-2019 & Madrid-2019  & Madrid-2019 \\

Ca$^{2+}$     &        &        &        &   &   &    &    &   &  &  &  Madrid-2019  & Madrid-2019   & Madrid-2019  & Madrid-2019   \\

O$_{w}$&        &        &        &        &   &   &    &    &   &  &    &      Madrid-2019    & Madrid-2019  & Madrid-2019  \\

S      &        &        &        &        &   &   &    &    &   &  &    &          & Madrid-2019  & Madrid-2019  \\

O$_{s}$&        &        &        &        &   &   &    &    &   &  &    &          &  & Madrid-2019 \\

\hline
\hline
    \end{tabular}}
  \end{center}
\end{table*}

 \begin{table*}[!hbt] \centering
\caption{Lennard-Jones parameters $\epsilon_{ij}$  (in kJ/mol)
for the ions Li$^{+}$, Na$^{+}$, K$^{+}$, Rb$^{+}$, Cs$^{+}$,
	 Mg$^{2+}$, Ca$^{2+}$, F$^{-}$, Cl$^{-}$, Br$^{-}$ I$^{-}$ and SO$_{4}^{2-}$.  O$_{w}$ and O$_{s}$ are the water an sulfate oxygens.}
\label{tab_eps}
  \begin{center}
 \resizebox{18cm}{!}{
    \begin{tabular}{c c c c c c c c c c c c c c c}
\hline
\hline
     & F$^{-}$ & Cl$^{-}$ & Br$^{-}$ & I$^{-}$  & Li$^{+}$ & Na$^{+}$ & K$^{+}$ & Rb$^{+}$ & Cs$^{+}$     & Mg$^{2+}$     & Ca$^{2+}$      &O$_{w}$ & S & O$_{s}$ \\

\hline
F$^{-}$ & 0.0309637 & LB &    LB &    LB &    0.1102655 &     LB &    0.223167 &      0.2161202 &     0.097105 &      LB &    LB &    0.1000 & LB & LB\\

Cl$^{-}$     & &      Madrid-2019     & LB &  LB &    Madrid-2019 &   Madrid-2019 &   Madrid-2019 & 0.340641 &        0.1615558 &     Madrid-2019     & Madrid-2019 & Madrid-2019 & Madrid-2019 & Madrid-2019\\

Br$^{-}$     &        &        &  0.112795 &  LB      & 0.199378 &    0.35677 & 0.425940 &    0.458323 &      0.195632 &      0.641807 &      0.239185 &      0.1000 & LB & LB \\

I$^{-}$     &        &        &        &     0.17901  & 0.273498      & 0.513387      & 0.536590 &    0.519646 &      0.246452 &      0.808534 &      0.301320 &      0.1000  & LB & LB \\

Li$^{+}$     &  &    &    &     & Madrid-2019 &       Madrid-2019     & Madrid-2019 & LB &    LB &    Madrid-2019 &   Madrid-2019 &   Madrid-2019 & Madrid-2019 & Madrid-2019\\

Na$^{+}$     &        &  &   &   &  & Madrid-2019     & Madrid-2019 & LB &    LB &    Madrid-2019     & Madrid-2019 & Madrid-2019 & Madrid-2019 & Madrid-2019 \\

K$^{+}$      &        &        & &     &    &  & Madrid-2019 &        LB &    LB      & Madrid-2019 & Madrid-2019 &   Madrid-2019 & Madrid-2019 & Madrid-2019\\

Rb$^{+}$      &        &        &  &     &    &   &   & 1.862314 &    LB &    LB &    LB &    0.1000 & LB & 1.250800\\

Cs$^{+}$      &        &        &    &     &     &     &    &    & 0.3759596 &        LB      & LB &  0.1000 & LB & LB\\

Mg$^{2+}$     &        &        &        &   &   &    &    &   &  & Madrid-2019 &      Madrid-2019     & Madrid-2019 & Madrid-2019 & Madrid-2019 \\

Ca$^{2+}$     &        &        &        &   &   &    &    &   &  &  &  Madrid-2019  & Madrid-2019   & Madrid-2019 & Madrid-2019  \\

O$_{w}$&        &        &        &        &   &   &    &    &   &  &    &      Madrid-2019   & Madrid-2019 & Madrid-2019  \\

S      &        &        &        &        &   &   &    &    &   &  &    &   &   Madrid-2019    & Madrid-2019  \\

O$_{s}$ &        &        &        &        &   &   &    &    &   &  &    &         &  & Madrid-2019 \\

\hline
\hline
    \end{tabular}}
  \end{center}
\end{table*}

In Table \ref{tabla-solubilidades} we show experimental melting temperatures  and
solubility limits for the salts considered in this work. As can be seen, with the exception of LiF 
and NaF which have low solubilities, the rest of the salts have medium and high solubilities.
In the case of fluorides (KF, RbF and CsF), the solubility is extremely high, even reaching a
solubility of 37 m for CsF. The solubility of LiBr, used in energy conversion processes is also high.

\begin{table}[!hbt]
\caption{Experimental melting temperature for anhydrous salt\cite{haynes2014} and salt solubility
in water\cite{haynes2014} at 25 $^{\circ}$C reported in molality units for the 
 salts studied in this work.}
\label{tabla-solubilidades}
  \begin{center}
    \begin{tabular}{ c c c c c c c c c c c }
\hline
\hline
Salt & & Melting Temperature  & & Solubility at 25 $^{\circ}$C  \\
  && K & & mol/kg  \\
\hline
LiF & & 1121.35 & & 0.052\\
LiBr & & 825.15 & & 20.84\\
LiI & & 742.15 & & 12.33\\
NaF & & 1269.15 & & 0.99\\
NaBr & & 1020.15 & & 9.20\\
NaI & & 933.15 & & 12.4\\
KF & & 1131.15 & & 17.50\\
KBr & & 1007.15 & & 5.77\\
KI & & 954.15 & & 8.92\\
RbF & & 1106.15 & & 28.8\\
RbCl & & 988.15 & & 7.77\\
RbBr & & 955.15 & & 7.01\\
RbI & & 915.15 & & 7.76\\
Rb$_{2}$SO$_{4}$ & &1323.15 & & 1.90\\
CsF & & 976.15 & & 37.7\\
CsCl & & 918.15 & & 11.3\\
CsBr & & 909.15 & & 5.77\\
CsI & & 894.15 & & 3.26\\
Cs$_{2}$SO$_{4}$ & & 1278.15 & & 5.03\\
MgBr$_2$ & & 984.15 & & 5.6\\
MgI$_2$ & & 907.15 & & 5.2\\
CaBr$_2$ & & 1015.15 & & 7.65\\
CaI$_2$ & & 1056.15 & & 7.3\\
\hline
\hline
    \end{tabular}
  \end{center}
\end{table}

\section{Simulation details}

We have studied different properties to test our extended model.
Molecular Dynamics (MD) simulations have been carried out using 
the GROMACS package \cite{spoel05,hess08} in the $NpT$ and $NVT$ ensembles.
In all the runs the leap-frog integrator algorithm\cite{bee:jcp76} with a time step of 2 fs was used.
We also applied  periodic boundary conditions in all directions in all cases.
The temperature was kept constant using the Nos\'e-Hoover thermostat\cite{nose84,hoover85} 
with a coupling constant of 2 ps. Parrinello-Rahman barostat\cite{parrinello81} 
with time constant of 2 ps
was implemented to keep constant pressure in $NpT$ simulations (1 bar for all the simulations).
For electrostatics and van der Waals interactions the cut-off radii were fixed at 1.0 nm and long-range 
corrections to the Lennard-Jones part of the potential in the energy and pressure were applied. 
The smooth PME method \cite{essmann95} to account for
the long-range electrostatic forces was used.
Water geometry was mantained using
the LINCS algorithm\cite{hess97,hess08b}.

 Most of the results of this work (unless otherwise stated) for aqueous solutions were obtained from $NpT$ simulations using 
 555 molecules of water and from runs lasting 50 ns.  Densities, water diffusion coefficients and radial distribution 
 functions were obtained from these runs. The choice of the number of water molecules is 
useful because for 1:1 electrolytes, adding ten cations and ten anions would yield a solution being approximately 
1 m (i.e 1 mol of salt per kilogram of water). A property of interest is the number of CIP which 
represents the number of anions that are in close contact with a cation (i.e without a molecule of water between 
the two ions). The number of CIP is evaluated easily from the cation-anion pair correlation function as:

\begin{equation}
    n^{CIP}= 4\pi \rho_{\pm}\int_{0}^{r_{min}} g_{\pm}(r)\;r^{2}\;dr,
\label{eq_cip}
\end{equation}

where $g_{\pm}$ is the cation-anion radial distribution function (RDF) and $\rho_{\pm}$ is the number density
of cation or anions (number of cations or anions per unit of volume), $r_{min}$ (the integral
upper limit)  is the
position of the first minimum in the RDF  which must be located at a
similar distance to that of the cation-O$_{w}$ RDF. One can  plot
simultaneously the RDFs  cation-anion and cation-O$_{w}$ to determine if we are
really evaluating the CIP or a contact solvent separated ion pair (CSSIP which corresponds to a 
cation in contact with an anion but with a molecule of water between them).
The hydration numbers (i.e the number of water molecules
around each ion) can be also calculated in a similar equation
to Eq.~\eqref{eq_cip} but
replacing $\rho_{\pm}$ by $\rho_{w}$ (i.e the number density of water obtained by dividing
the number of molecules of water between the volume of the simulation box)
and $g_{\pm}(r)$ by g$_{ion-O_{w}}(r)$ instead.

The Einstein relation was used to calculate diffusion coefficients:
\begin{equation}
\label{eq_diffus}
  D = \lim_{t \to \infty} \frac{1}{6 t}  \Big \langle [\Vec{r}_{i}(t)-\Vec{r}_{i}(t_{0})]^{2} \Big \rangle ,
\end{equation}
where $\Vec{r}_{i}(t)$ and $\Vec{r}_{i}(t_{0})$ are the position of the i$^{th}$
particle at time $t$ and a certain origin of time $t_{0}$ and the $\langle
[r_{i}(t)-r_{i}(t_{0})]^{2}\rangle$ term is the mean square displacement (MSD).
From the plot of the MSD versus time a slope can be obtained which is 6 times the diffusion coefficient.

 A somewhat larger system having 4440 molecules of water (i.e eight times 555) was considered 
 to evaluate viscosities and the possible existence of precipitation. 
The methodology used to compute the viscosity is similar to that described in previous
works\cite{gon:jcp10}.
We perform a  previous $NpT$ simulation to
calculate accurately the volume of the system. After that, a
$NVT$ simulation of 50 ns was performed.
Throughout the run, the pressure tensor $P_{\alpha\beta}$
was calculated and saved on disk every 2 fs.
The
off-diagonal elements of the pressure tensor, are
	$P_{xy}$, $P_{xz}$ and
$P_{zy}$ which are equivalent. Besides, due to the rotational
invariance
of the molecules, the terms ($P_{xx}$ - $P_{yy}$)/2 and ($P_{yy}$ - $P_{zz}$)/2
are also equivalent\cite{guo01,alfe98}. Thus, we have averaged the five pressure components
in order to obtain accurate results.
Finally, the Green-Kubo formula for the viscosity was used:
\begin{equation}
\label{shear_GK}
 \eta = \frac {V}{kT} \int_{0}^{\infty} 
 \langle P_{\alpha\beta}(t_{0})\; P_{\alpha\beta}(t_{0}+t) \rangle_{t_{0}}\; dt.
\end{equation}
The  upper limit of the integral is usually between 10-20 ps.

Aggregation and precipitation will spoil the results of any simulation giving unphysical results. 
We have not determined the solubility of the salts considered in the force field of this work.
However we have performed a simple test. We have performed a long run of 50 ns at the experimental 
value of the solubility limit of each salt, using a large system having 4440 molecules of water (since 
nucleation time decreases with system size we used a large system to be on the safe side) 
and with the number of ions required to mimic the experimental value of the solubility and checked 
for the absence of precipitation. The absence of precipitation was checked in several ways such as analyzing 
visually the trajectories, analyzing the final configuration of the run, from a dynamic increase of the number of 
CIP, and from the possible existence of a drift in the thermodynamic properties. 
For all the salts considered in this work 
we found no evidence of precipitation at the experimental value of the solubility limit. 
Of course this does not guarantee that the force field in this work has the correct solubility.
It only guarantees that the salt is either stable or metastable at the experimental value of the solubility 
limit and that the presented results are not an artefact due to the existence of spontaneous precipitation. 

Molten densities were
obtained for systems containing 1000 ions. Simulations typically lasted
50~ns in a $NpT$ simulation at 1 bar and at the melting temperature of the salt.
We usually performed
simulations with integer charges first (with densities close to the experimental values),
and then used the final configuration as the initial one for 
$NpT$ runs using scaled charges to observe the decrease in density provoked by the use of scaled charges. 
The average densities were obtained from the last 20 ns of the run, after the system was fully relaxed.

\section{Results}

\subsection{Finding parameters for Rb$^{+}$ and Cs$^{+}$: Chloride salts}

 We shall start by presenting results for chloride salts as they 
 were obtained to determine parameters for Rb$^{+}$ and Cs$^{+}$
 (i.e Rb$^{+}$-Rb$^{+}$ , Cs$^{+}$-Cs$^{+}$, Rb$^{+}$-Cl$^{-}$, Cs$^{+}$-Cl$^{-}$
 and more importantly Rb$^{+}$-water and Cs$^{+}$-water interactions).
 When developing the Madrid-2019-Extended force field we typically used the following 
 strategy. We considered a salt (or several) with formula XY, 
 either X (or Y) being an ion not present in the Madrid-2019 force field and Y (or X) an ion for 
 which parameters are available in the Madrid-2019 force field. 
 For instance, to determine parameters for Rb$^{+}$  we shall use as target properties 
 those of the RbCl salt. We used this salt to obtain the Rb$^{+}$-water interaction.
 As mentioned before Rb$^{+}$-Cl$^{-}$ and Rb$^{+}$-Rb$^{+}$
 interactions were determined by fitting the properties of the melt 
 and keeping the number of CIP within reasonable limits.  We follow
 a similar procedure to determine the parameters of Cs$^{+}$.

\begin{center}
\begin{figure}[!hbt] \centering
    \centering
    \includegraphics*[clip,scale=0.3,angle=0.0]{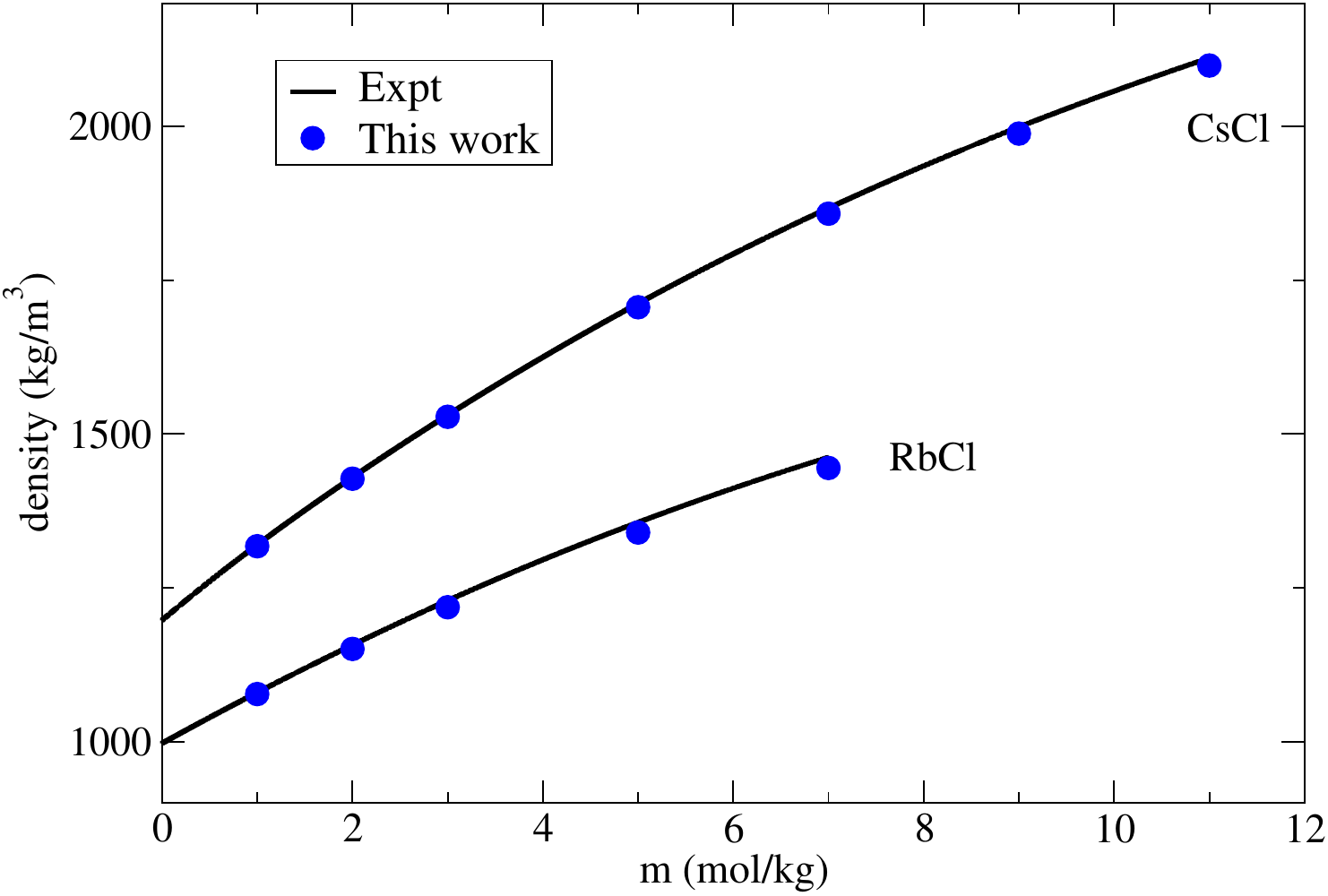}
    \caption{Density as a function of molality at T = 298.15 K 
    and 1 bar for chloride salts aqueous solutions, RbCl and CsCl. 
    Blue circles: this work. Solid black  lines: fit of experimental data taken from 
    Ref\cite{tables}.
CsCl values were shifted up 200 density units for better legibility.}
    \label{density-rbcl-cscl}
\end{figure}
\end{center}

   In Figure \ref{density-rbcl-cscl} the densities from experiments are compared to the simulation 
   results obtained from the Madrid-2019-Extended force field for concentrations up to the experimental 
   value of the solubility limit. As can be seen the agreement is quite good. Only at high concentrations
   are the experimental values slightly underestimated.
   The statistical error in the densities calculated in this work is always less than 0.25\%.

\begin{center}
\begin{figure}[!hbt] \centering
    \centering
    \includegraphics*[clip,scale=0.3,angle=0.0]{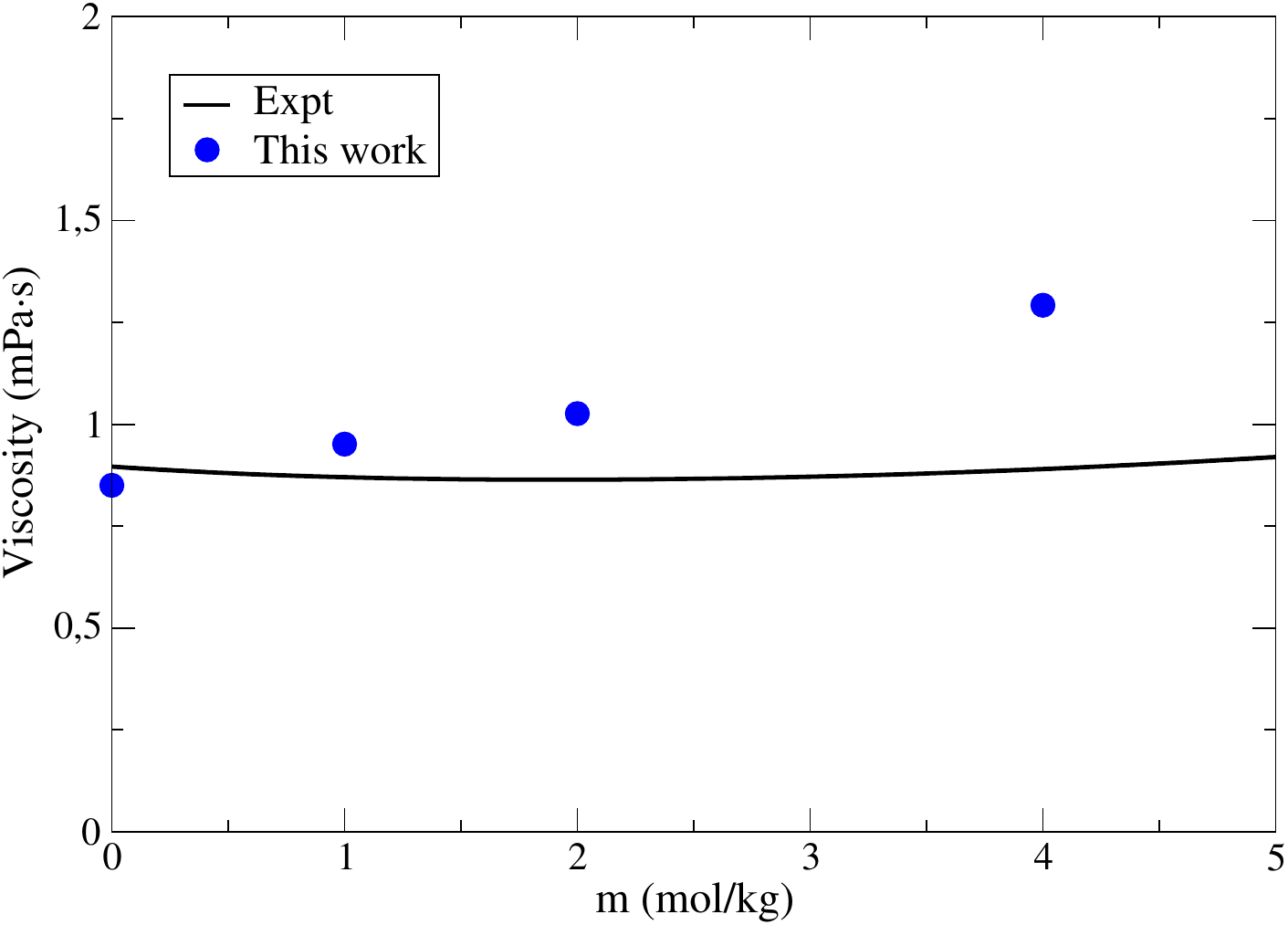}
    \caption{Shear viscosity curves as a function of concentration for aqueous RbCl
    solutions at 298.15 K and 1 bar. Blue circles are the results from this work.
    The continuous lines are the fit of experimental data taken from Ref\cite{ostroff}. 
    }
    \label{visco-rbcl}
\end{figure}
\end{center}

  Next we computed the viscosity. Results are shown in Figure \ref{visco-rbcl}.
  The statistical error in the viscosities calculated in this work is always less than 5-10\%
  (being lower at low concentrations of salt).
  Experimentally the viscosity of a RbCl solution does not change much with concentration, 
  decreasing slightly at low concentrations, having a weak minimum and then increasing 
  slightly. The simulations are not able to capture this subtle behavior. 
  Also it is clear that the model overestimates the viscosity with respect to experimental values. However 
  the model is able to predict that the change in the viscosity of a RbCl solution 
  with respect to water is much smaller than that of a NaCl or KF solution at similar 
  concentrations as will be discussed later in this paper. Further work is needed to understand 
  the origin of this discrepancy. We checked for the possible existence of spontaneous precipitation 
  in the simulations. However we found no evidence of precipitation at the highest concentration 
  considered, so that this is not at the origin of the discrepancy. 

 Structural properties are presented in Table \ref{tab_res_cloruros}. 
Before describing the structural properties of the following chloride salts, 
it is important to point out that the experimental data with which we compare were measured 
in general, at low concentrations and for this reason
they depend only on the ion being studied 
and not on the particular salt.
Regarding specifically chloride salts, it can be seen that the hydration 
 number of the cation at high concentration is around 6. The number of CIP is below
0.5 for both salts. For RbCl the number of CIP is 0.23 at 7 m and for CsCl is 0.48
for 11 m. Thus these cations have around 6.3 (Rb$^{+}$) or 6.6 (Cs$^{+}$) particles around them. 
At high concentrations ions can replace water molecules (although due to the different sizes the 
replacement is not one to one).
The distance at which first the peak appears for the cation-oxygen radial distribution function is 
slightly smaller than that found in experiments. However densities are predicted well
as it can be seen in Figure \ref{density-rbcl-cscl}.

\begin{table*}[!hbt]
\caption{Structural properties for chloride electrolyte solutions at 298.15 K 
and 1 bar.  Number of contact ions pairs (CIP), hydration number of cations
(HN$_{c}$) and anions (HN$_{a}$), and position of the first maximum of the
cation-water ($d_{c-O_{w}}$), and anion-water ($d_{a-O_{w}}$) in the radial
distribution function. In
parentheses, experimental data taken from the work of Marcus\cite{mar:cr88}.
Properties were calculated at low concentrations and close to the solubility limit of each salt.}
\label{tab_res_cloruros}
  \begin{center}
    \begin{tabular}{ c c c c c c c c}
\hline
\hline
	    Salt   & $m$  & CIP   & HN$_{c}$    & HN$_{a}$   & $d_{c-O_{w}}$ & $d_{a-O_{w}}$ \\
	      &(mol/kg) &    &     &    &  \AA& \AA\\
\hline
	    RbCl   & 1 & 0.05  & 6.3(5-8)     & 5.8(5.3-7.2)     & 2.75(2.79-2.90) & 3.04(3.08-3.34) \\
	    & 7 & 0.23  & 6.0     & 5.6     & 2.75 & 3.04 \\
	    CsCl  & 1  & 0.07  & 6.8(8-9)     & 5.9(5.3-7.2)     & 2.87(2.95-3.20) & 3.04(3.08-3.34) \\
	    & 11  & 0.48  & 6.0   & 5.5   & 2.85 & 3.04 \\
\hline
\hline
    \end{tabular}
  \end{center}
\end{table*}

\subsection{Fluoride salts}

\begin{center}
\begin{figure}[!hbt] \centering
    \centering
    \includegraphics*[clip,scale=0.3,angle=0.0]{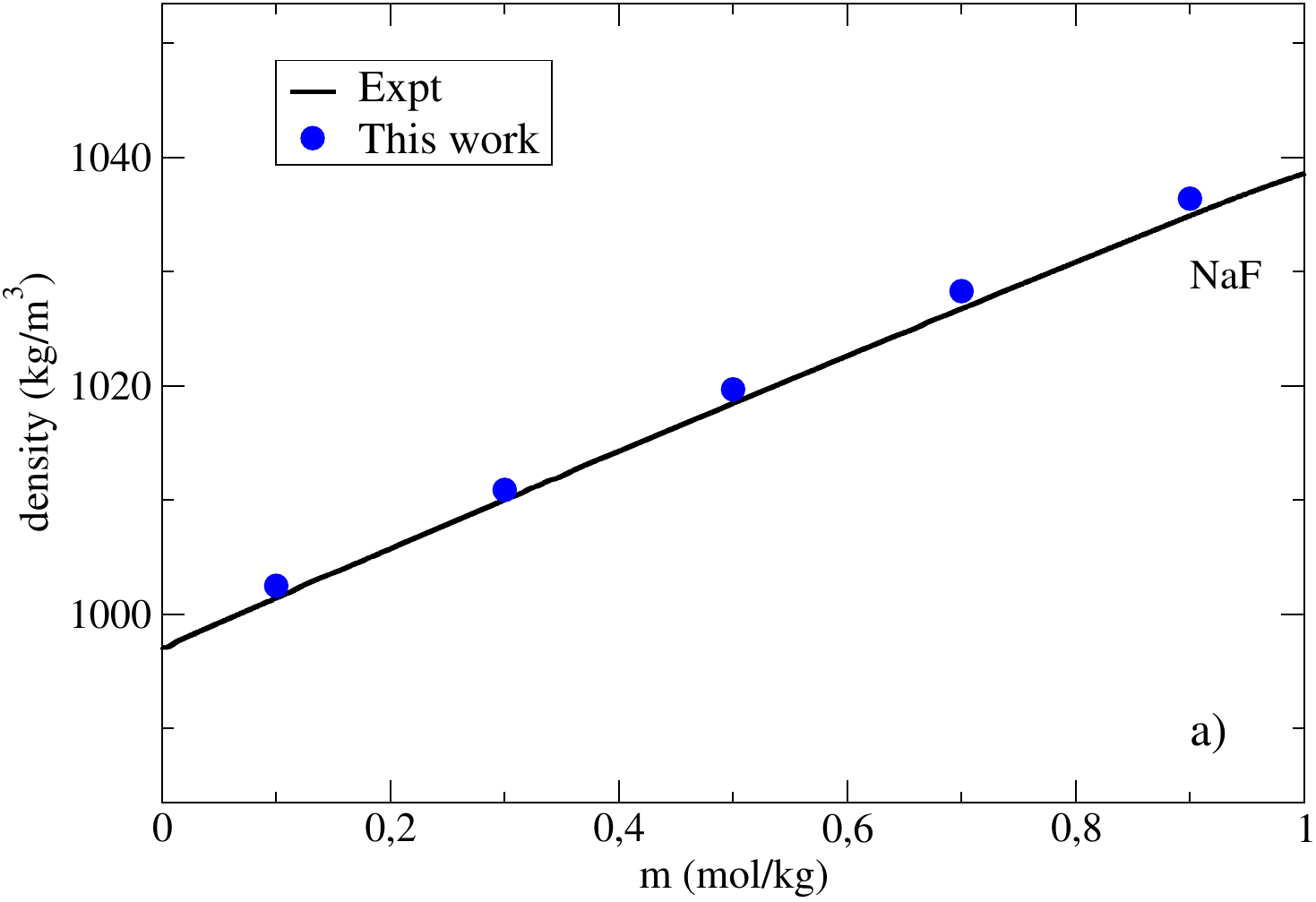}
    \includegraphics*[clip,scale=0.3,angle=0.0]{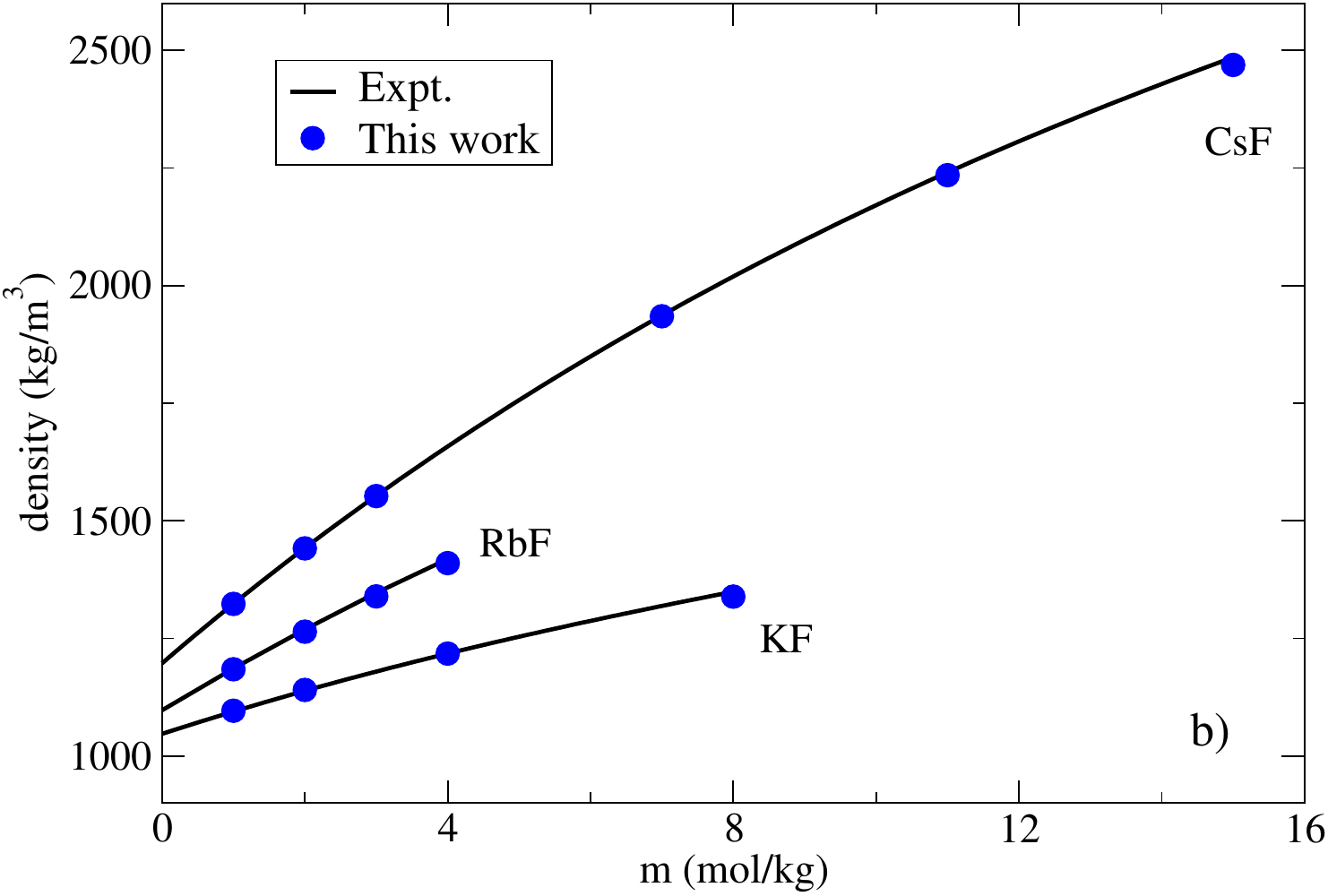}
    \caption{Density as a function of molality at T = 298.15 K 
	and 1 bar, blue circles (this work), solid black lines (fit of experimental data taken 
	from Ref.\cite{CRC_1984_49_445} for NaF and KF, Ref.\cite{tables} for RbF and
    Ref.\cite{doi:10.1021/je00051a018} for CsF. a)  NaF aqueous solutions b)  KF, RbF and CsF aqueous solutions. 
       RbF densities were shifted up 100  and CsF 200 density units for a clear visualization.}
    \label{density-fluoruros}
\end{figure}
\end{center}

The densities of different fluoride salts have been calculated with the 
Madrid-2019-Extended force field. 
A comparison with experimental results is shown in Figure \ref{density-fluoruros}. 
Since the solubility of NaF is small, the results for this salt are presented separately in  Figure \ref{density-fluoruros}a).
(LiF was not considered as its solubility is extremely small, i.e 0.052 m). As can be seen the 
agreement with experiment for NaF is quite good. 
The results for KF, RbF and CsF are presented in  Figure \ref{density-fluoruros}b).
The solubilities  of KF, RbF and CsF are extremely high and increase with the size of the cation. 
Good agreement with experiment is also found for these salts. For KF, RbF and CsF we did not find experimental 
results at high concentrations (even though we have evaluated the densities of these salts
in the whole range of concentrations up to their solubility
limit. These results are collected in the supplementary material of this work). Given the accuracy of the simulations at low-moderate concentrations, 
one could expect that simulations results at high concentrations should be reasonable.

\begin{center}
\begin{figure}[!hbt] \centering
    \centering
    \includegraphics*[clip,scale=0.3,angle=0.0]{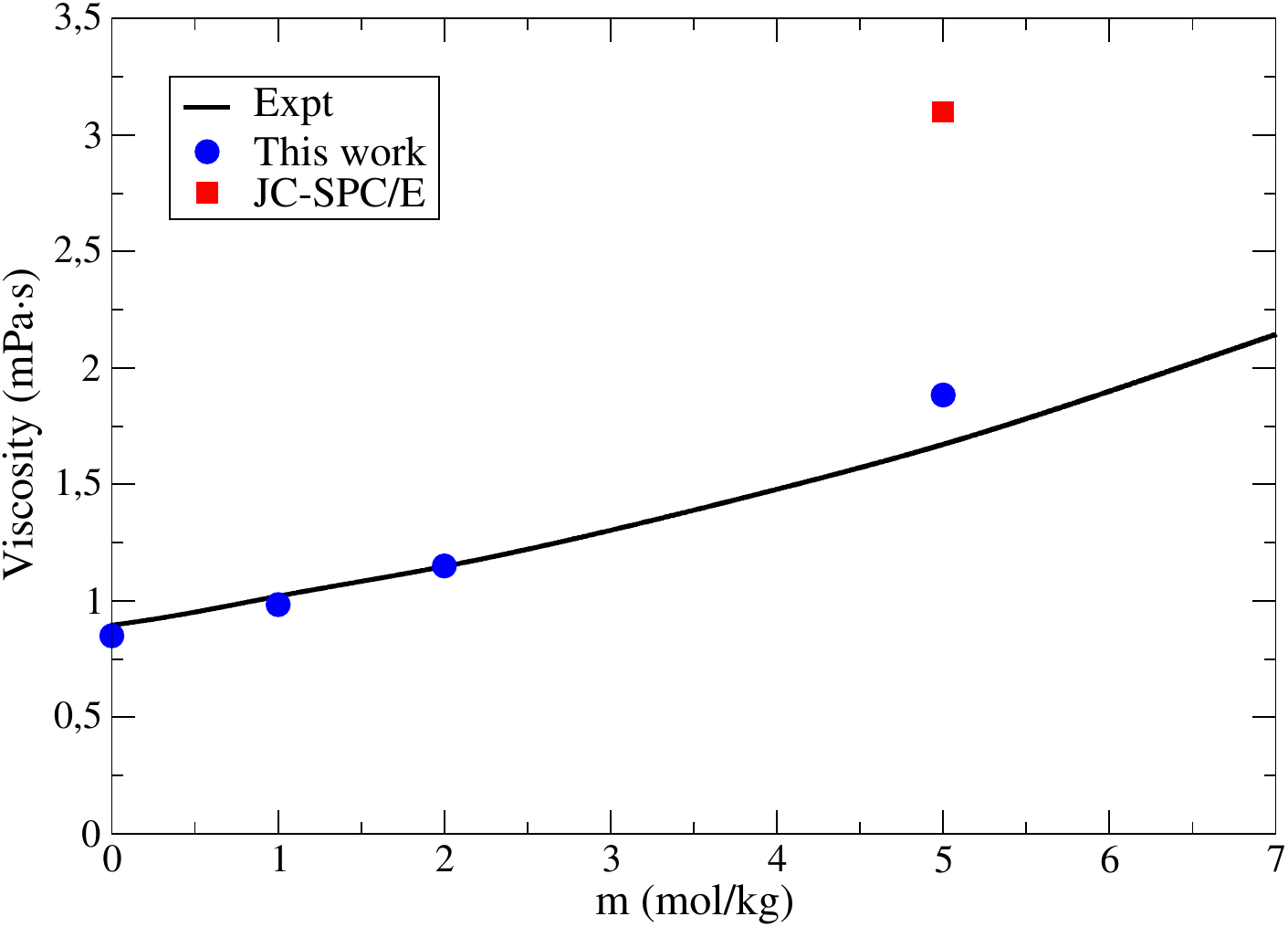}
    \caption{Shear viscosity curves as a function of concentration for aqueous KF
    solutions at 298.15 K and 1 bar. Blue circles are the results from this work. Red squares
        are the results for JC-SPC/E model and
    the continuous black line is the fit of experimental data taken from Ref\cite{CJC_56_1442_1978}.}
    \label{visco-kf}
\end{figure}
\end{center}

\begin{center}
\begin{figure}[!hbt] \centering
    \centering
    \includegraphics*[clip,scale=0.3,angle=0.0]{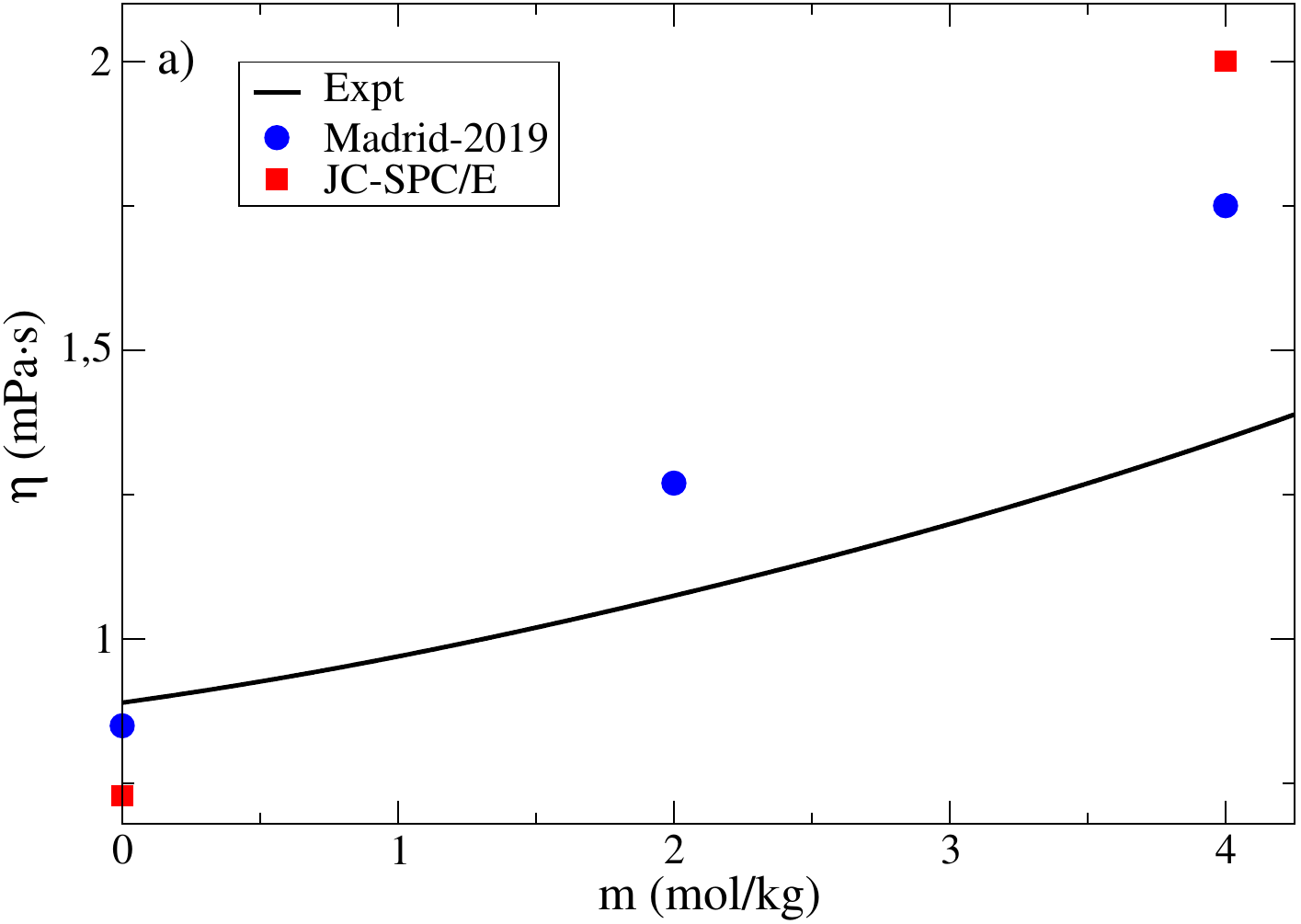}
    \includegraphics*[clip,scale=0.3,angle=0.0]{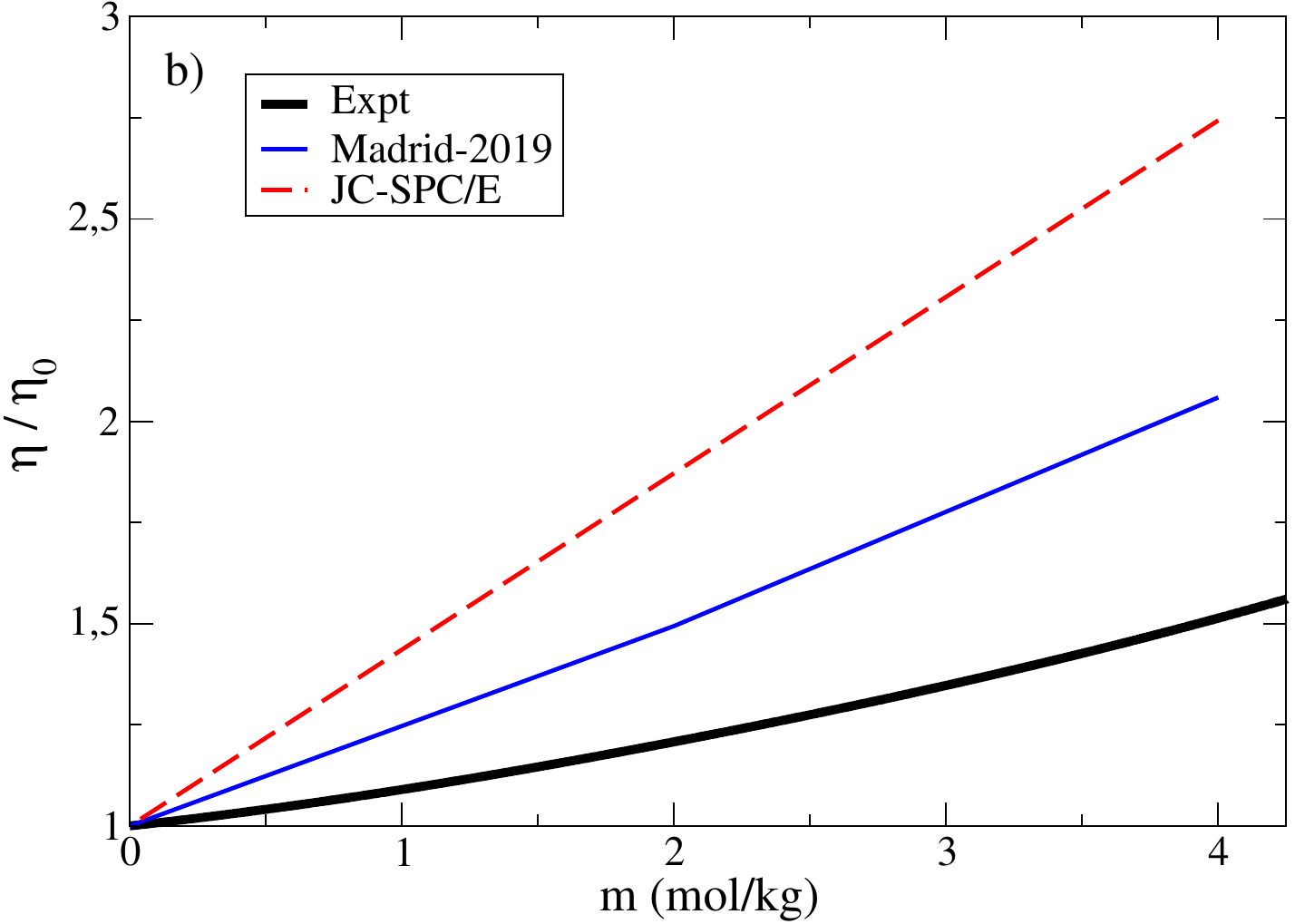}
    \caption{Viscosity as a function of concentration for NaCl 
	solutions at 298.15 K and 1 bar. a) Shear viscosity results: Blue circles are the results from this work with original
	Madrid-2019 force field. Red squares
        are the results for JC-SPC/E model and
    the continuous black line is the fit of experimental data taken from ref\cite{lal:jced07,doi:10.1021/je700232s}.
        b) Ratio of the  viscosity at different concentrations to the the viscosity in pure water. The blue solid line 
	represents our data for Madrid-2019 model, red dashed line results for JC-SPC/E model and black solid thick line is the fit of the
	experimental data.}
    \label{visco-nacl}
\end{figure}
\end{center}

 Let us now present some results for the viscosities. Since the evaluation of the viscosity is rather 
 expensive in this work we shall evaluate the viscosity only for some selected salts. 
 In Figure \ref{visco-kf} the results for the viscosity of the KF are presented. 
 It can be observed that the Madrid-2019-Extended force field of this work predicts quite well the viscosities for 
 concentrations up to 2 m, and reasonably well for the most concentrated 5 m solution. 
 The model overestimates somewhat the experimental value. 
 This behavior is similar to the one found for the salts included in the original Madrid-2019 force field\cite{JCP_2019_151_134504}. 
 In Figure \ref{visco-kf}  we also present the viscosity of the JC-SPC/E model (which uses integer charges) 
 as determined in this work. 
 It is clear that in this case the viscosity is overestimated at 5 m by a factor of two. 
 Thus for KF it seems that the use of scaled charges improves the description of the viscosity. 
 To analyze if the overestimate of the viscosity of KF by the JC-SPC/E is an exception, we have 
 also computed the viscosity of the JC-SPC/E for NaCl, which is arguably the most important salt. Somewhat 
 surprisingly its value at room temperature and pressure and at high concentrations has not been reported before (to the best of our 
 knowledge). The value of the viscosity of NaCl, both from the original Madrid-2019 and from the JC-SPC/E
 model are presented in Figure \ref{visco-nacl}a). 
 Again it is clear that the JC-SPC/E overestimates the experimental value of the viscosity 
 of NaCl solutions at high concentrations. This is not a problem of the water model as in our previous 
 work we showed  that also the JC model of NaCl overestimates the viscosity even when used with 
 the TIP4P/2005 model of water\cite{JCP_2019_151_134504}. Thus, at least for NaCl and KF it is clear that the JC-SPC/E model overestimates
 the value of the viscosity. This overestimation is even greater if we evaluate the ratio of the model viscosity at different concentrations
 to the model viscosity in pure water as we can see in Figure \ref{visco-nacl}b).  This is in line with the results presented by
 Yue and Panagiotopoulos\cite{doi:10.1080/00268976.2019.1645901}. At low concentrations it was shown that the viscosity of the JC-SPC/E increases
 faster than found in experiments, and that scaled and polarizable models of NaCl exhibited better (although
 not perfect) agreement with experiment.

To complete the study of the fluoride salts with Madrid-2019-Extended
force field we have analyzed some
structural results. 
We have calculated the cation-anion, cation-water and anion-water
radial distribution functions close to the experimental
solubility limit of each salt.

\begin{table*}[!hbt]
\caption{Structural properties for fluoride electrolyte solutions at 298.15 K 
(291.15 K for RbF)
and 1 bar.  Number of contact ions pairs (CIP), hydration number of cations
(HN$_{c}$) and anions (HN$_{a}$), and position of the first maximum of the
cation-water ($d_{c-O_{w}}$), and anion-water ($d_{a-O_{w}}$) in the radial
distribution function. In
parentheses, experimental data taken from the work of Marcus\cite{mar:cr88}.
Properties were calculated close to the solubility limit of each salt and at low concentrations.}
\label{tab_res_fluoruros}
  \begin{center}
    \begin{tabular}{ c c c c c c c c}
\hline
\hline
Salt  & $m$ & CIP   & HN$_{c}$    & HN$_{a}$   & $d_{c-O_{w}}$ & $d_{a-O_{w}}$ \\
              &(mol/kg) &    &     &    &  \AA& \AA\\
\hline
	    NaF  & 0.1  & 0.00  & 5.5(4-8)   & 5.8(6-9)   & 2.33(2.40-2.50) & 2.77(2.54-2.87) \\
	         & 0.9  & 0.02  & 5.5   & 5.8   & 2.33 & 2.77 \\
	    KF    & 1 & 0.04  & 5.7(6-8)     & 5.7(6-9)   & 2.73(2.60-2.80) & 2.76(2.54-2.87) \\
         	  & 17 & 1.15  & 5.8     & 4.75   & 2.73 & 2.75 \\
	    RbF   & 1 & 0.06  & 6.4(5-8)     & 5.7(6-9)     & 2.76(2.79-2.90) & 2.77(2.54-2.87) \\
	          & 28 & 2.45  & 4.8     & 3.45     & 2.76 & 2.74 \\
	    CsF   & 1 & 0.05  & 6.9(8-9)     & 5.6(6-9)     & 2.86(2.95-3.20) & 2.76(2.54-2.87) \\
	          & 37 & 3.30  & 4.0     & 2.85    & 2.86 & 2.73 \\

\hline
\hline
    \end{tabular}
  \end{center}
\end{table*}

In Table \ref{tab_res_fluoruros} we have collected all the results obtained
for these structural properties and we have compared them with 
experimental X-ray and neutron diffraction data collected in the work of
Marcus\cite{mar:cr88}.
As with the anion-water distances  $d_{F^{-}-O_{w}}$  we see the value found in this work (around
   2.75~\AA)  is within 
  the experimental reported values 2.54-2.87~\AA. 
  With respect to the cation-water distance the values
  found in the simulations
  are in general slightly lower (except for K$^{+}$) than the lower bound reported in experiments.
  Similar behavior was found in the Madrid-2019 for the Li$^{+}$ cation.
  We do not have an explanation for this. In any case it seems that the prediction of the distance
  if wrong does not result in bad predictions for the densities.
  It can be seen that the $d_{cation-O_{w}}$ increases with the size of the cation.
  The difference between $d_{K^{+}-O_{w}}$ and $d_{Rb^{+}-O_{w}}$ distances are
  small  but as expected the $d_{Rb^{+}-O_{w}}$ is larger.
With respect to the hydration number one can see that the F$^{-}$ anion can be hydrated by about
5.7 molecules of water (see the results for NaF which is a system without CIP). For the other
salts it holds that the hydration number of the F$^{-}$ anion plus the number of CIP is around six.
Thus for each molecule of water removed from the hydration shell, a cation occupies its place indicating
that the size of both water and K$^{+}$, Rb$^{+}$ and Cs$^{+}$ cations although not identical are not so different.
We have to point out that
(with the exception of NaF) the rest of fluorides have very high solubility values.
Moreover, in the case of RbF, at the solubility limit 
the total number of ions is practically equal
to the number of water molecules and 
in the case of CsF it is even higher. Thus, we can not apply the same rules
as for other salts and it is clear that one should allow higher number of CIP.
It is also interesting to evaluate what a random mixing model would predict for the number of CIP.
If differences in size between water and the cations are neglected, then one would expect that the number of CIP
would be 6 $\cdot$ (17/55) = 1.85, 6 $\cdot$ (28/55)=3.05 and 6 $\cdot$ (37/55)=4.03 for KF, RbF and CsF respectively.
Six is the number of water molecules that can be located around the F$^{-}$ anion, and this is multiplied
by the ratio of the molality to the number of moles of water in one kilogram.
As can be seen the random mixing model would predict a number of CIP higher than found in the
simulations (1.15, 2.45 and 3.30 respectively for KF, RbF and CsF).
To be on the safe side we checked that no spontaneous precipitation occurred at the experimental
value of the solubility limit after 50 ns using a large system with 4440 molecules of water. 

  We shall now present results for the bromide salts.

\subsection{Bromide salts}

 \begin{center}
\begin{figure}[!hbt] \centering
    \centering
    \includegraphics*[clip,scale=0.3,angle=0.0]{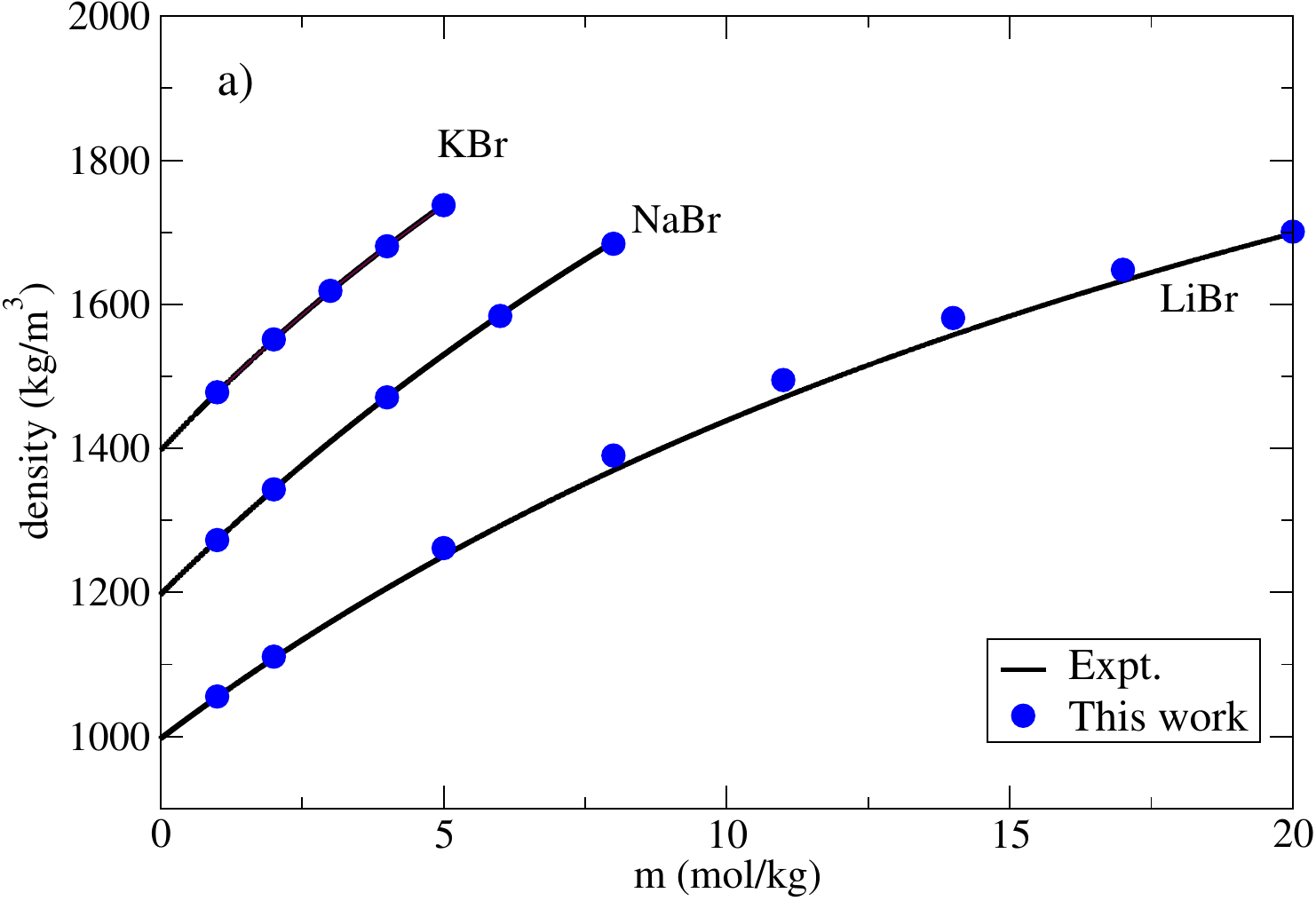}
    \includegraphics*[clip,scale=0.3,angle=0.0]{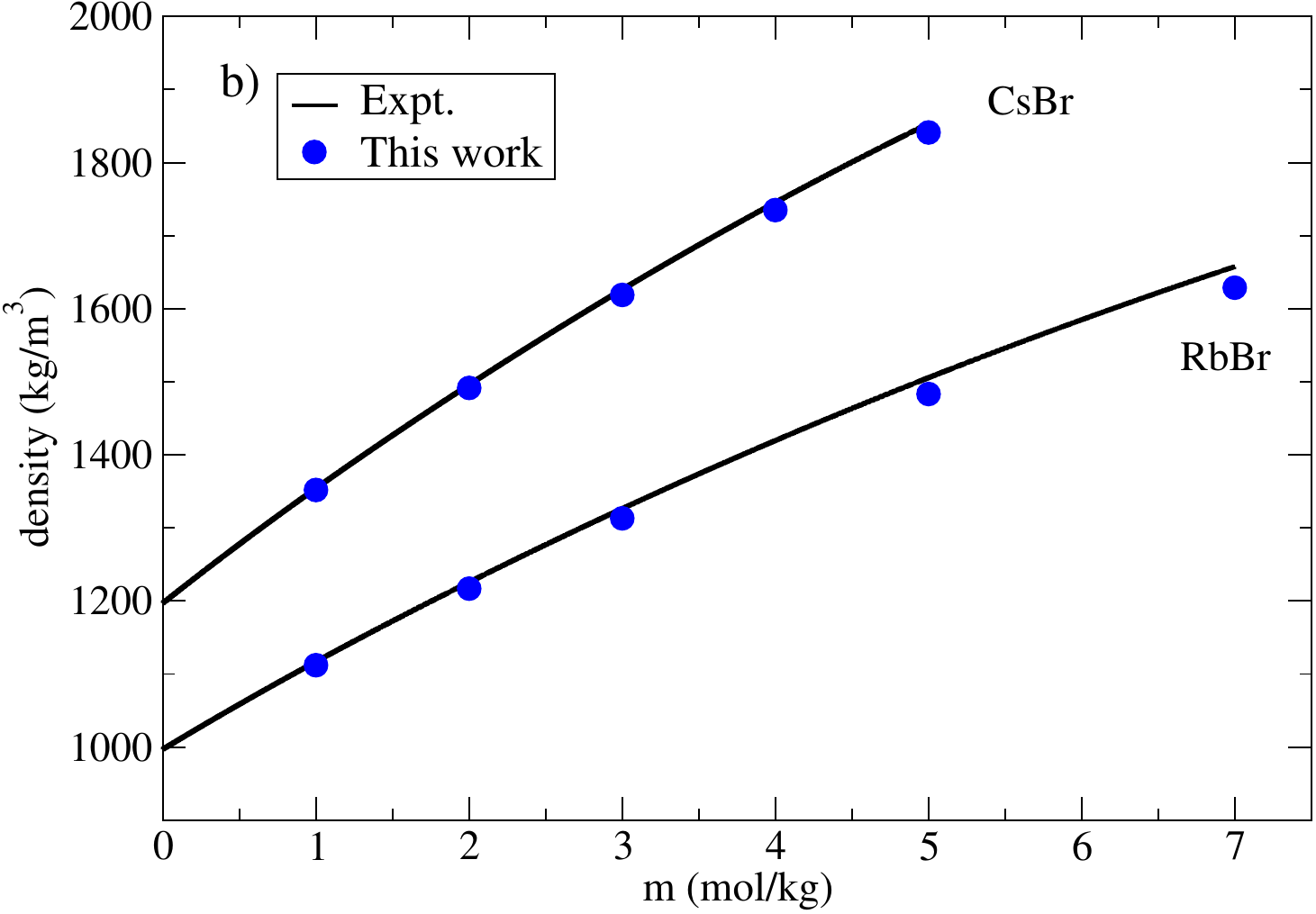}
    \caption{Density as a function of molality at T = 298.15 K 
	and 1 bar for bromide salts aqueous solutions. a) LiBr, NaBr and KBr. b) 
	RbBr and CsBr. 
    Blue circles: this work. Solid black  lines: fit of experimental data taken from 
    Ref\cite{tables} for all salts  and 
    Ref\cite{doi:10.1021/je00035a016} for NaBr and KBr.
NaBr and CsBr densities were shifted up 200 units  and KBr 400 density units for a clear visualization.}
    \label{density-bromuros}
\end{figure}
\end{center}

In Figure \ref{density-bromuros}a) we show the results for densities of 
LiBr, NaBr and KBr. 
The solubilities  of NaBr and KBr are moderate. The densities obtained for these 
salts are in excellent agreement with all the experimental data over all the
molality range. In the case of LiBr, the simulations results overestimate slightly
the experimental ones at intermediate molalities. Even though, the simulations results
are very accurate.
In Figure \ref{density-bromuros}b) results for the densities of RbBr and CsBr salt solutions are presented. 
As can be seen the results of Madrid-2019-Extended are in reasonable agreement with 
the experimental ones. It seems that the series of the bromides is challenging. 
In general the agreement found is good but sometimes the density is overestimated (as in LiBr) 
and sometimes is underestimated (as in RbBr and CsBr).

The number of CIP have been evaluated for these salts and are presented
in Table \ref{tab_res_bromuros}. NaBr at 8 m has a 
number of CIP of 0.24 and KBr at 5 m has a number of CIP of 0.29. Both of them follow
the rule proposed by Benavides et \textit{al.}\cite{ben:mp17} with a number of CIP
around or below
0.5. The same is true for RbBr and CsBr with a number of CIP of 0.58 and 0.37 respectively. 
Thus for these salts with solubility smaller than 10 m, the number of CIP is either around or 
below 0.5 at the solubility limit. However for LiBr, with a solubility of 20 m the number of 
CIP is 1.30.

\begin{center}
\begin{figure}[!hbt] \centering
    \centering
    \includegraphics*[clip,scale=0.3,angle=0.0]{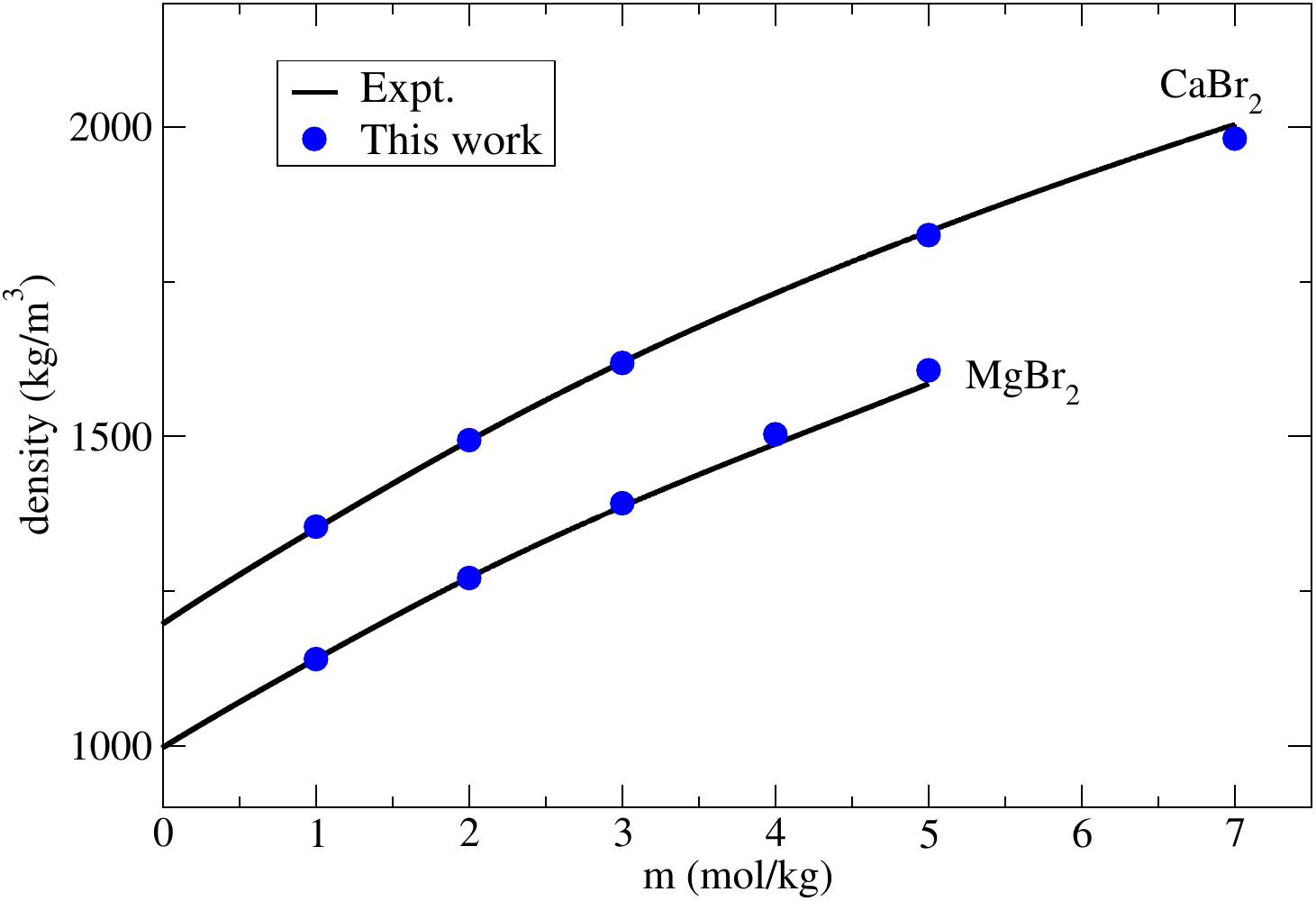}
    \caption{Density as a function of molality at T = 298.15 K 
    and 1 bar for bromide salts 1:2 aqueous solutions, MgBr$_2$ and CaBr$_2$. 
    Blue circles: this work. Solid black  lines: fit of experimental data taken from 
    Ref\cite{tables} and Ref\cite{doi:10.1021/je00051a018}.
 CaBr$_2$ values were shifted up 200 density units for better legibility.}
    \label{density-mgbr-cabr}
\end{figure}
\end{center}

We have also studied divalent salts for bromides. The original Madrid-2019 force field included parameters 
to describe the interaction between  Mg$^{2+}$ with water and between Ca$^{2+}$ with water.
Parameters for the  Mg$^{2+}$-Mg$^{2+}$  and  Ca$^{2+}$-Ca$^{2+}$ were also provided in our previous 
work. The Br$^{-}$-Br$^{-}$ interaction was obtained in this work from the study of 1:1 electrolytes containing 
Br$^{-}$. Thus for these salts we can only modify the Br$^{-}$-Mg$^{2+}$ and Br$^{-}$-Ca$^{2+}$ interactions. 
As can be seen in Figure \ref{density-mgbr-cabr}, up to  3 m the results of the force field reproduce the
experimental data. At high concentrations the results are reasonable for the model but tend to underestimate 
slightly the experimental results in the case of CaBr$_2$ and overestimate slightly them for MgBr$_2$. 
Regarding the number of contact ion pairs, for the
MgBr$_{2}$ is 0 and for CaBr$_{2}$ is 0.04. Both salts have low number of CIP and do not precipitate.
In this respect bromides behave like chlorides. Both Mg$^{2+}$  and Ca$^{2+}$ have a strong interaction with water,
and cation-anion contact are rare.

\begin{center}
\begin{figure}[!hbt] \centering
    \centering
    \includegraphics*[clip,scale=0.3,angle=0.0]{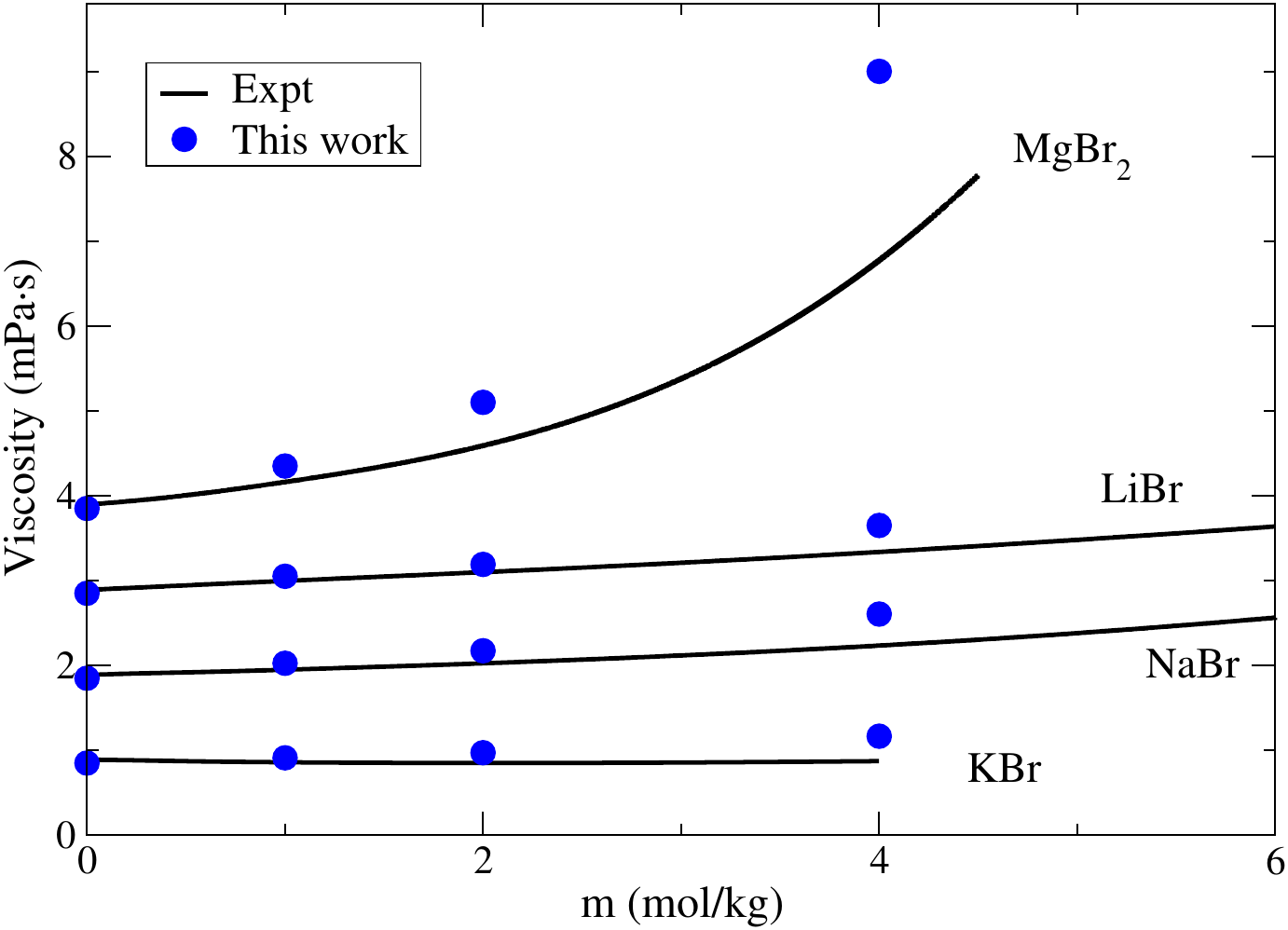}
	\caption{Shear viscosity curves as a function of concentration for aqueous bromides
    solutions at 298.15 K and 1 bar. Blue circles are the results from this work.
    The continuous lines are the fit of experimental data taken from Refs\cite{CJC_56_1442_1978,zitsev,doi:10.1021/j100470a024}. 
	NaBr, LiBr and MgBr$_2$ values were shifted up 1, 2 and 3 viscosity units respectively 
	for better legibility.}
    \label{visco-bromides}
\end{figure}
\end{center}

 Let us now study the behavior of the viscosities for the bromide salts.
 In particular we shall analyze LiBr, NaBr, KBr and MgBr$_2$. Results are shown in 
 Figure \ref{visco-bromides}. It is interesting that experimentally, LiBr,
 NaBr and KBr have a small impact on the
 viscosity of water (increasing it slightly in the case of NaBr and LiBr) and decreasing 
 it slightly in the case of KBr. The Madrid-2019-Extended force field is able to 
 capture this effect. Although the agreement with experiment is not quantitative
 the trends are described quite well.  However adding MgBr$_2$ significantly increases  
 the viscosity of water. Again this effect is captured by the force field but  it seems
 that the model overestimates the magnitude of the effect (and the deviation is already visible 
 at 2 m). This is similar to the 
 behavior found for MgCl$_2$. For these salts, MgCl$_2$ and MgBr$_2$, the scaled charges do a good 
 job in describing the experimental values of the viscosities but tend to overestimate 
 their values. 

  We shall now present the results for the structural properties. 
The hydration number of the bromide anion is around 6. Except for LiBr, for the 
rest of salts the sum of the hydration number and the number of CIP is around 6.
That makes sense when taking into account that Na$^{+}$, K$^{+}$, Rb$^{+}$ and Cs$^{+}$
have sizes similar to that 
of water. However the exception is LiBr. For this salt, at 20 m one has 7.5 molecules around
the Br$^{-}$ anion, 6.3 water and 1.5 Li$^{+}$. Thus Li$^{+}$ when in the first coordination of Br$^{-}$, 
provokes a contraction of the molecules of water hydrating the bromide anion.
The bromide-water distance found in this work is within the range of values reported 
in experiments. For the cations again, the cation-water distance found in the simulation 
is always slightly below the value reported in experiments. We do not have an explanation 
for this finding, especially taking into account that the predictions of the densities 
for bromide salts is quite reasonable. 

\begin{table*}[!hbt]
\caption{Structural properties for bromide electrolyte solutions at 298.15 K 
and 1 bar.  Number of contact ions pairs (CIP), hydration number of cations
(HN$_{c}$) and anions (HN$_{a}$), and position of the first maximum of the
cation-water ($d_{c-O_{w}}$), and anion-water ($d_{a-O_{w}}$) in the radial
distribution function. In
parentheses, experimental data taken from the work of Marcus\cite{mar:cr88}.
Properties were calculated at low concentrations and close to the solubility limit of each salt.}
\label{tab_res_bromuros}
  \begin{center}
    \begin{tabular}{ c c c c c c c c}
\hline
\hline
	    Salt   &$m$ & CIP   & HN$_{c}$    & HN$_{a}$   & $d_{c-O_{w}}$ & $d_{a-O_{w}}$ \\
              &(mol/kg) &    &     &    &  \AA& \AA\\
\hline
	    LiBr  & 1  & 0.18  & 3.8(3.3-5.3)   & 6.3(4-6)   & 1.84(1.90-2.25) & 3.15(3.01-3.45) \\
	    & 20  &  1.5   &  2.0  &  5.6   &  1.84 & 3.15 \\
	    NaBr  & 1 & 0.02  & 5.5(4-8)     & 6.0(4-6)   & 2.34(2.40-2.50) & 3.15(3.01-3.45) \\
	    & 8 & 0.24  & 5.3     & 6.0   & 2.33 & 3.15 \\
	    KBr    & 1 & 0.04  & 6.7(6-8)     & 5.9(4-6)   & 2.73(2.60-2.80) & 3.15(3.01-3.45) \\
	    & 5 & 0.29  & 6.4     & 5.8  & 2.73 & 3.15 \\
	    RbBr   & 1 & 0.10  & 6.3(5-8)     & 5.8(4-6)     & 2.76(2.79-2.90) & 3.15(3.01-3.45) \\
	    & 7 & 0.58  & 5.8     & 5.4    & 2.75 & 3.15 \\
	    CsBr   & 1 & 0.08  & 6.7(8-9)     & 5.9(4-6)     & 2.85(2.95-3.20) & 3.15(3.01-3.45) \\
	    & 5 & 0.37  & 6.2     & 5.6     & 2.85 & 3.15 \\
	    MgBr$_2$   & 1 & 0.00  & 6.0(6-8.1)     & 5.9(4-6)     & 1.92(2.00-2.11) & 3.15(3.01-3.45) \\
	    & 5 & 0.00  & 6.0     & 5.9     & 1.92 & 3.15\\
	    CaBr$_2$   & 1 & 0.00  & 7.4(5.5-8.2)     & 6.1(4-6)     & 2.38(2.39-2.46) & 3.15(3.01-3.45) \\
	    & 7 & 0.04  & 6.9     & 6.1     & 2.38 & 3.15 \\
\hline
\hline
    \end{tabular}
  \end{center}
\end{table*}

\begin{center}
\begin{figure}[!hbt] \centering
    \centering
    \includegraphics*[clip,scale=0.3,angle=0.0]{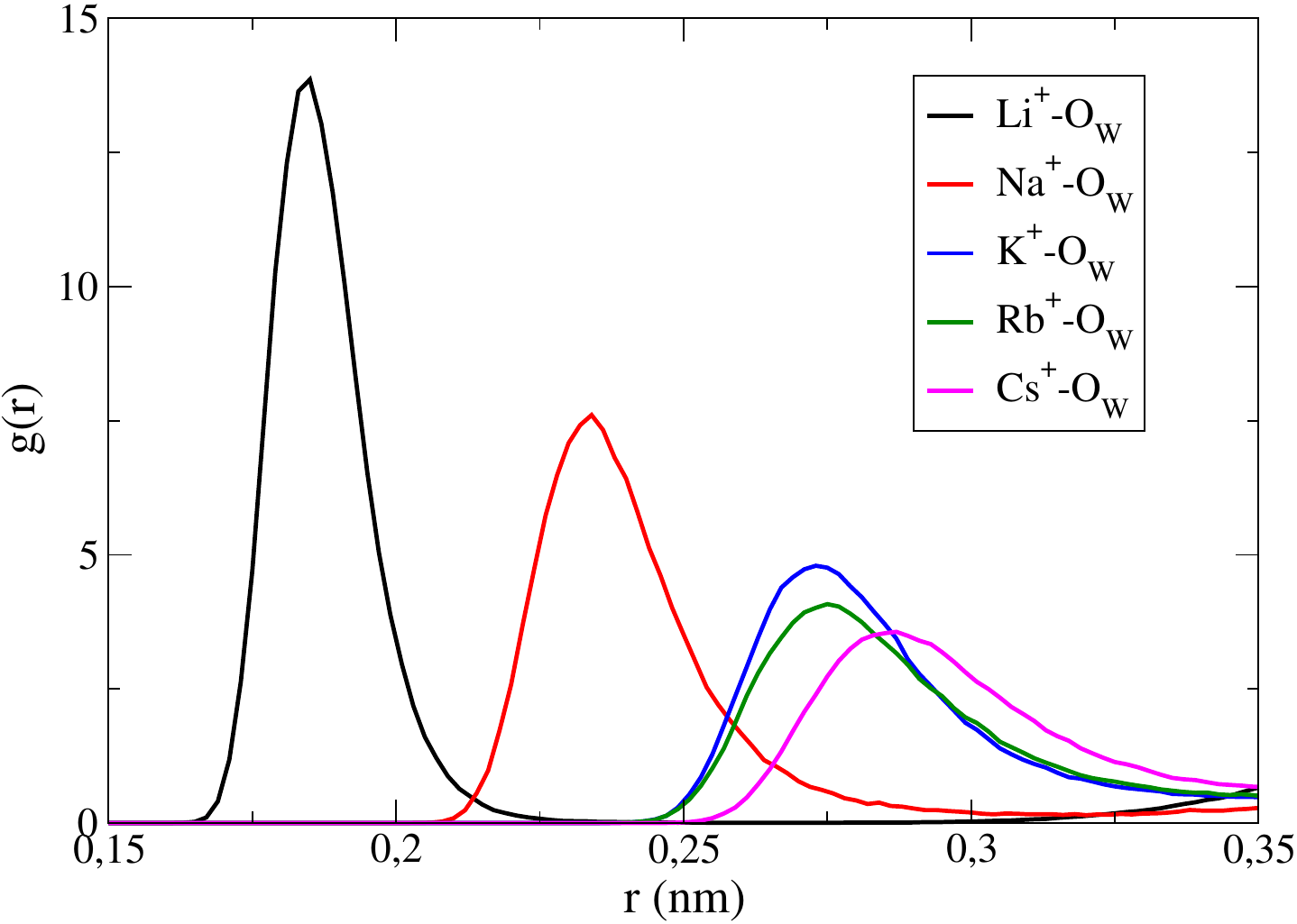}
    \caption{Cation-water oxygen radial distribution function for bromide
    solutions
at 298.15 K, 1 bar, and 1 m as were obtained with the Madrid-2019-Extended model in
solutions: LiBr, NaBr, KBr, RbBr and CsBr}
    \label{estructural-cationes}
\end{figure}
\end{center}

Finally in Figure \ref{estructural-cationes} we show the cation-water radial distribution function
as obtained in this work for a 
1 m solution of LiBr, NaBr, KBr, RbBr and CsBr. 
It can be seen that the distance at which the first peak occurs increases with the size of the cation. 
The increase is clear for all the cations with one exception. The K$^{+}$-O$_{w}$  and Rb$^{+}$-O$_{w}$ 
have a similar location of the first peak (i.e 2.73 vs 2.75 \AA \; respectively).

\subsection{Iodine salts}

\begin{center}
\begin{figure}[!hbt] \centering
    \centering
    \includegraphics*[clip,scale=0.3,angle=0.0]{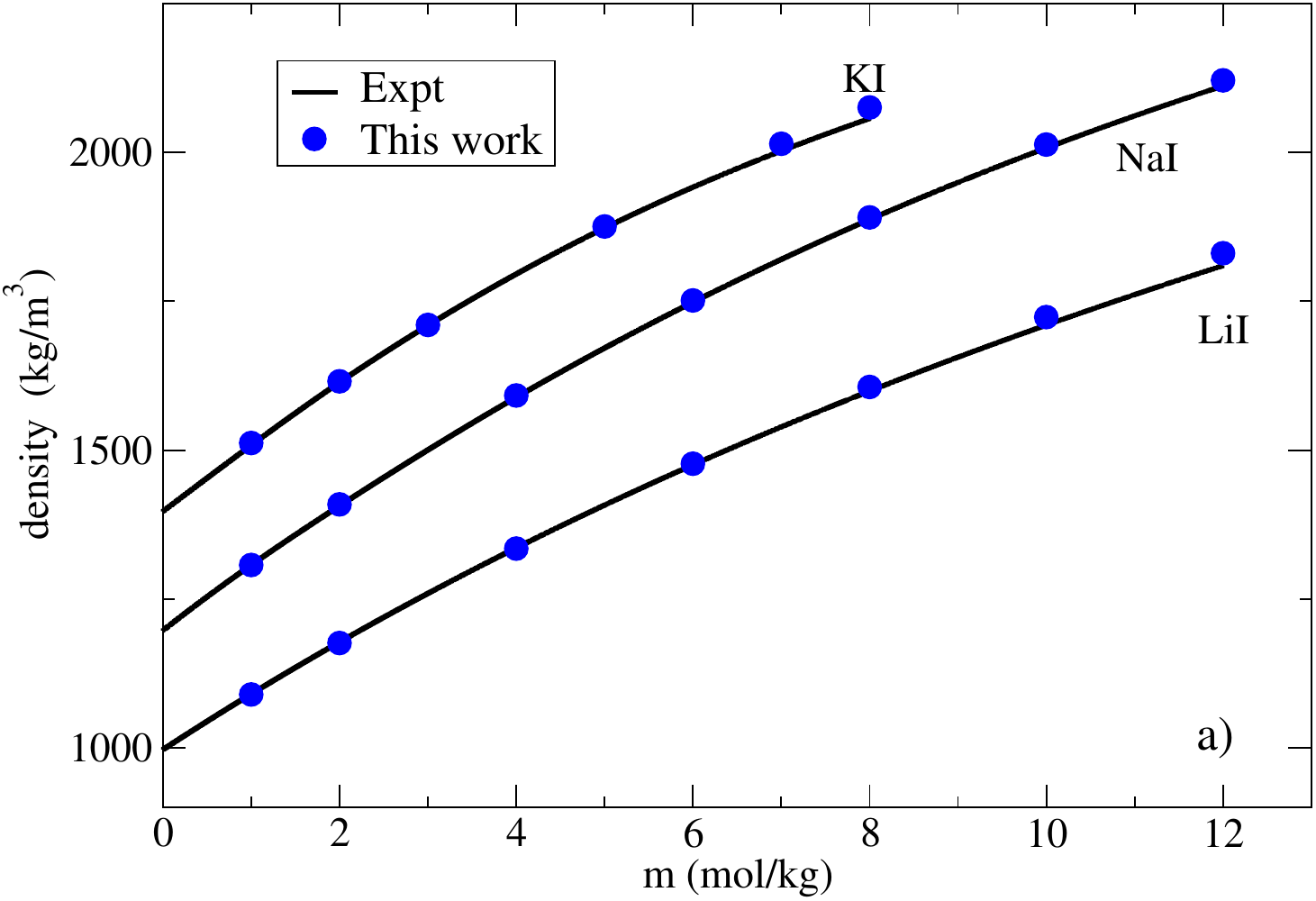}
    \includegraphics*[clip,scale=0.3,angle=0.0]{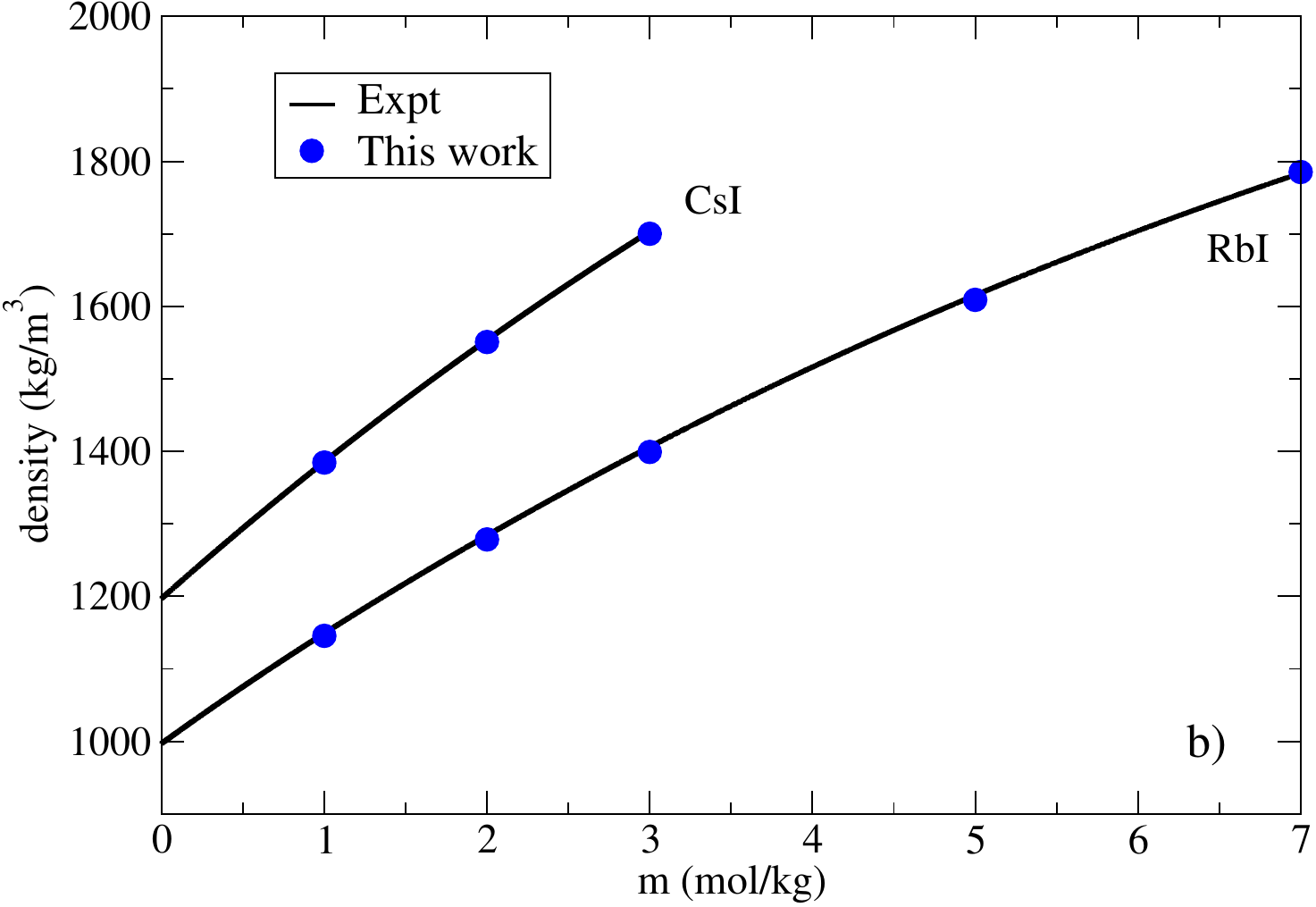}
    \caption{Density as a function of molality at T = 298.15 K 
	and 1 bar for iodide salts aqueous solutions. a) LiI, NaI and KI. b)
	and RbI and CsI.
    Blue circles: this work. Solid black  lines: fit of experimental data taken from 
    Ref\cite{tables}.
NaI and CsI densities were shifted up 200  and KI 400 density units for better legibility.}
    \label{density-yoduros}
\end{figure}
\end{center}

 We present now the properties for the iodine salts. In particular we shall consider LiI, NaI, KI, RbI, CsI, MgI$_2$ and 
 CaI$_2$. Thus for iodine salts we present results for a number of cations, including monovalent and divalent cations. 
We shall start by presenting densities for monovalent ions as obtained from the Madrid-2019-Extended. 
Results for LiI, NaI and KI are shown in Figure \ref{density-yoduros}a), and results for RbI and CsI in Figure \ref{density-yoduros}b). 

LiI, NaI and KI simulation results are in excellent agreement with 
the experimental ones for the whole range of molalities. 
 However, at the highest molalities studied for each salt (8 m for KI and 12 m for NaI and LiI) the simulation 
 results overestimate slightly the experimental values. 
The results for RbI and CsI aqueous solutions are shown in Figure \ref{density-yoduros}b).
The results for both salts are in excellent agreement with experiment for all concentrations.

\begin{center}
\begin{figure}[!hbt] \centering
    \centering
    \includegraphics*[clip,scale=0.3,angle=0.0]{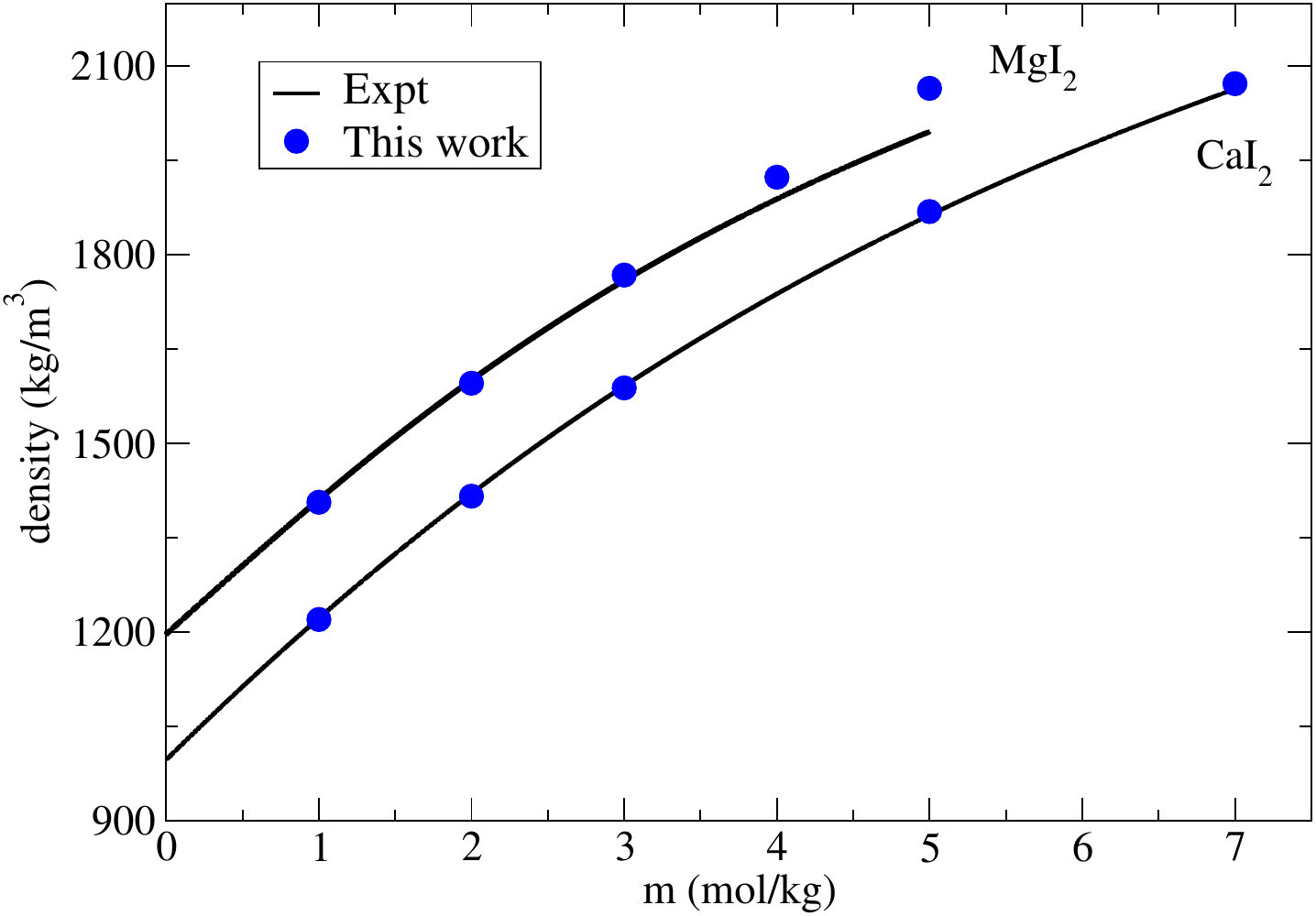}
    \caption{Density as a function of molality at T = 298.15 K 
    and 1 bar for iodide salts 1:2 aqueous solutions, MgI$_2$ and CaI$_2$. 
    Blue circles: this work. Solid black  lines: fit of experimental data taken from 
    Ref\cite{tables}.
 MgI$_2$ values were shifted up 200 units for better legibility.}
    \label{density-mgi-cai}
\end{figure}
\end{center}

We shall now turn to salts containing divalent cations. 
Figure \ref{density-mgi-cai} shows the results for the densities of divalent salts
of iodide. Results for CaI$_2$ are in excellent agreement with experimental data over all
the molality range. On the other hand, results for MgI$_2$ overestimate the experimental
values at high molalities.

\begin{center}
\begin{figure}[!hbt] \centering
    \centering
    \includegraphics*[clip,scale=0.3,angle=0.0]{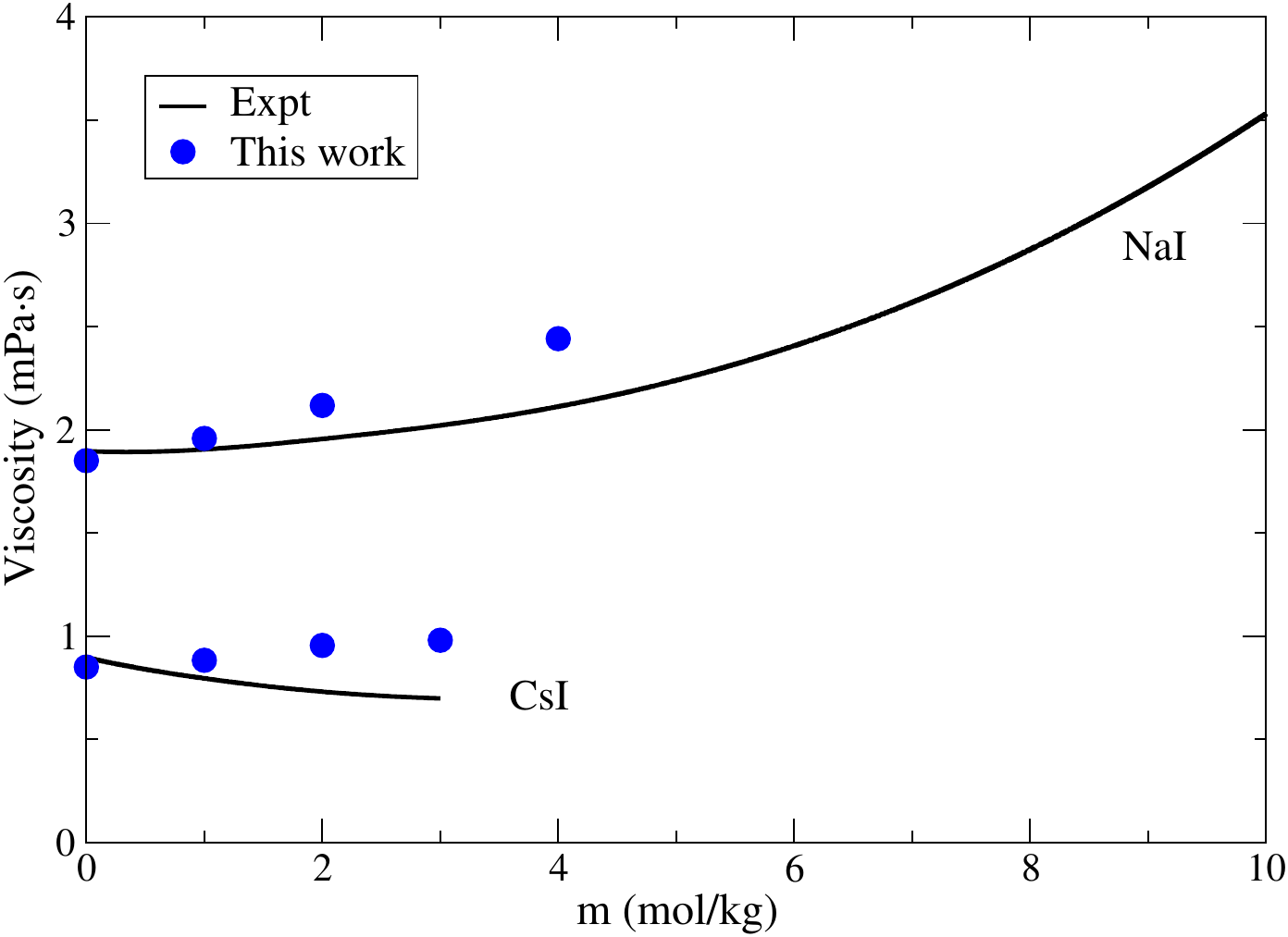}
    \caption{Shear viscosity curves as a function of concentration for aqueous NaI and
    CsI
    solutions at 298.15 K and 1 bar. Blue circles are the results from this work.
    The continuous lines are the fit of experimental data taken from
    Refs\cite{doi:10.1021/je060124c,doi:10.1021/je700232s} for NaI and Ref\cite{doi:10.1021/ja01295a026} 
    for CsI. NaI values were shifted up 1 viscosity unit for better legibility.
    }
    \label{visco-yoduros}
\end{figure}
\end{center}

Figure \ref{visco-yoduros} shows the viscosities obtained with Madrid-2019-Extended force field for NaI and CsI.
The agreement is not perfect, but still reasonable. 
The case of CsI is special because in experiments  the viscosity 
decreases slightly  as the salt concentration increases. 
The model is not able to capture this decrease in viscosity but at 
least the increase that occurs is very small.
The case of NaI is similar to the other salts studied in this work and in Madrid-2019\cite{JCP_2019_151_134504} 
force field. The viscosities are well predicted for concentrations up to 1-2 m and overestimated afterwards. 
The overestimation is not dramatic but seems to a systematic deviation found in the Madrid-2019-Extended force field. 
Very little is known about the behavior of the viscosities at high concentrations for most of the force fields of ionic systems. 
The study of Yue and Panagiotopoulos\cite{doi:10.1080/00268976.2019.1645901} and the results for KF and NaCl presented before seems to suggest that integer charges deviate 
typically more from experiment than scaled charges. It is also clear that scaled charges improve the description but are 
not able to obtain a full quantitative agreement with experiment. Notice that the deviations are not due to an incorrect prediction 
of densities. The Madrid-2019-Extended force field yields good prediction of the experimental densities. Therefore, the deviations 
are due to some missing physics in the model.  

Structural properties for iodine salts are presented in Table \ref{tab_res_yoduros}. 
The hydration number of iodine is about 6.1 (although this number changes from one salt 
to another at the high concentrations considered in Table \ref{tab_res_yoduros}).
The number of CIP is almost zero for Li$^{+}$, Mg$^{2+}$ and Ca$^{+}$, thus reflecting that these ions are 
strongly hydrated and CIP are rare. For the rest of the salts the number of CIP is 
typically below 0.5 at the solubility limit. The main exception to this rule is
the NaI with a higher number of CIP.

\begin{table*}[!hbt]
\caption{Structural properties for iodine electrolyte solutions at 298.15 K 
and 1 bar.  Number of contact ions pairs (CIP), hydration number of cations
(HN$_{c}$) and anions (HN$_{a}$), and position of the first maximum of the
cation-water ($d_{c-O_{w}}$), and anion-water ($d_{a-O_{w}}$) in the radial
distribution function. In
parentheses, experimental data taken from the work of Marcus\cite{mar:cr88}.
Properties were calculated at low concentrations and close to the solubility limit of each salt.
 }
\label{tab_res_yoduros}
  \begin{center}
    \begin{tabular}{ c c c c c c c c}
\hline
\hline
	    Salt   &$m$  & CIP   & HN$_{c}$    & HN$_{a}$   & d$_{c-O_{w}}$ & d$_{a-O_{w}}$ \\
	                  &(mol/kg) &    &     &    &  \AA& \AA\\
\hline
	    LiI   &1 & 0.00  & 4.0(3.3-5.3)   & 6.1(4-6)   & 1.84(1.90-2.25) & 3.28(3.01-3.45) \\
&12 & 0.01  & 4.0   & 6.1   & 1.84 & 3.28 \\
	    NaI   & 1 & 0.01  & 5.5(4-8)     & 6.1(4-6)   & 2.33(2.40-2.50) & 3.28(3.01-3.45) \\
	   & 12 & 1.12  & 5.3     & 6.1   & 2.33 & 3.28 \\
	    KI  & 1 & 0.03  & 6.5(6-8)     & 6.0(4-6)   & 2.72(2.60-2.80) & 3.29(3.01-3.45) \\
	    & 8 & 0.30  & 6.2     & 6.1   & 2.72 & 3.28  \\
	    RbI   & 1 & 0.10  & 6.3(5-8)     & 6.0(5.3-7.2)     & 2.75(2.79-2.90) & 3.29(3.08-3.34) \\
	    & 7 & 0.60  & 5.8     & 5.5     & 2.75 & 3.28 \\
	    CsI   & 1 & 0.10  & 6.6(8-9)     & 6.0(5.3-7.2)     & 2.86(2.95-3.20) & 3.28(3.08-3.34) \\
	    & 3 & 0.35  & 6.4     & 5.9     & 2.86 & 3.28 \\
	    MgI$_2$   &1  & 0.00  & 6.0(6-8.1)     & 6.1(4-6)     & 1.92(2.00-2.11) & 3.28(3.01-3.45) \\
	    &5  & 0.00  & 6.0     & 6.1     & 1.92 & 3.28 \\
	    CaI$_2$   & 1 & 0.00  & 7.4(5.5-8.2)     & 6.0(4-6)     & 2.38(2.39-2.46) & 3.28(3.01-3.45) \\
	    & 7 & 0.00  & 6.9     & 6.6     & 2.38 & 3.28 \\
\hline
\hline
    \end{tabular}
  \end{center}
\end{table*}

\begin{center}
\begin{figure}[!hbt] \centering
    \centering
    \includegraphics*[clip,scale=0.3,angle=0.0]{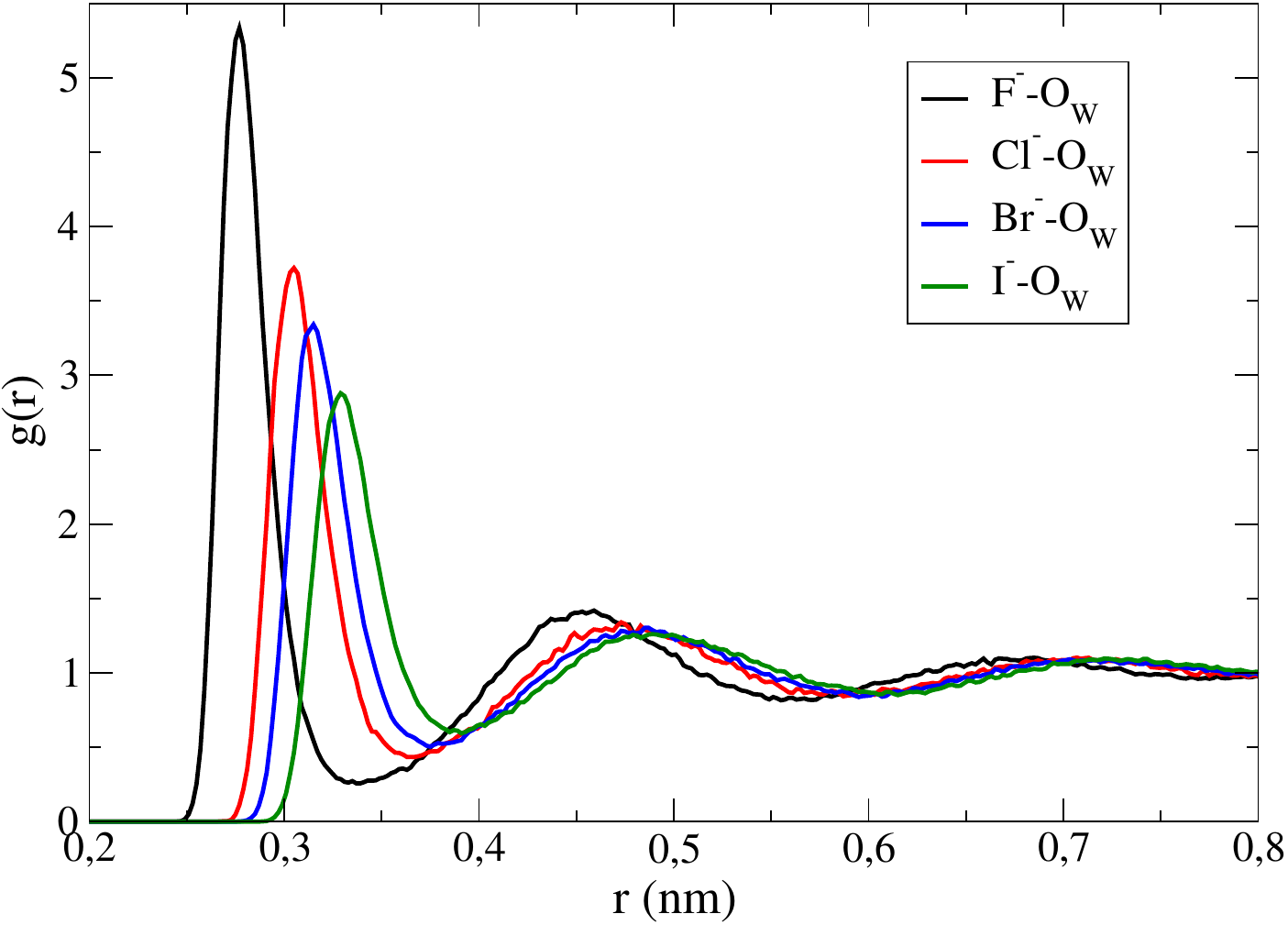}
    \caption{Anion-water oxygen radial distribution functions for sodium
    solutions
at 298.15 K, 1 bar, and 1 m as were obtained with the Madrid-2019-Extended model in
solutions: NaF (0.9 m), NaCl, NaBr and NaI.}
    \label{estructural-aniones}
\end{figure}
\end{center}

  To finish this section we shall present results for the anion-oxygen radial distribution function 
  for F$^{-}$, Cl$^{-}$, Br$^{-}$ and I$^{-}$. 
Figure \ref{estructural-aniones} shows the anion-water oxygen radial 
distribution functions for 1 m sodium salts solutions
with the different anions developed in 
Madrid-2019 and Madrid-2019-Extended force fields. It can be seen that the 
anion-oxygen distance increases with the size of the anion.

\subsection{
	Sulfate salts}

\begin{center}
\begin{figure}[!hbt] \centering
    \centering
    \includegraphics*[clip,scale=0.3,angle=0.0]{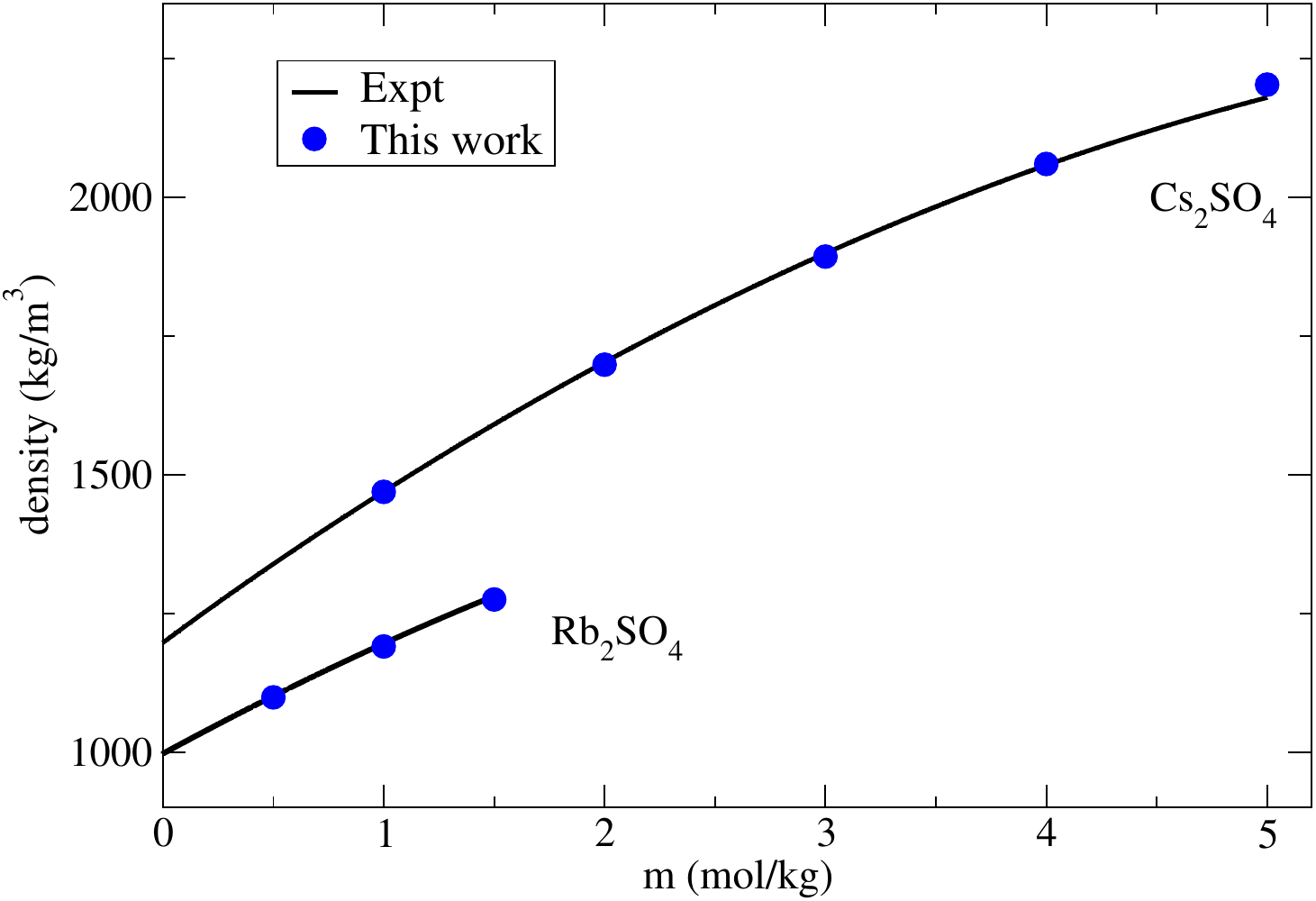}
	\caption{Density as a function of molality at T = 298.15 K
    and 1 bar for sulfate salts aqueous solutions, Rb$_2$SO$_4$ and
	 Cs$_2$SO$_4$.
    Blue circles: this work. Solid black  lines: fit of experimental data taken from
    Refs\cite{tables,densidades-cs2so4,densidades-cs2so4-chinos}.
	Cs$_2$SO$_4$ values were shifted up 200 density units for better legibility.}
    \label{density-sulfatos}
\end{figure}
\end{center}

We shall finish by presenting properties for the sulfate salts.
We have included the Rb$_2$SO$_4$ and
	 Cs$_2$SO$_4$. The results for the densities are in excellent agreement
	 with experimental results as we show in Figure \ref{density-sulfatos}.
	 Regarding the structural properties we can observe similar 
	 results than those obtained for the sulfate salts studied in the 
	 Madrid-2019 force field. The number of contact ion pairs is higher than
	 for other salts (sulfate group is a huge ion) and the hydration numbers
	 are higher than the experimental ones. In our opinion, the 
	 results obtained by using simulations are more realistic than
	 the ones from experiments because the size of the sulfate group
	 is too big to have only 8 molecules of water around.

\begin{table*}[!hbt]
\caption{
Structural properties for sulfate electrolyte solutions at 298.15 K
and 1 bar.  Number of contact ions pairs (CIP), hydration number of cations
(HN$_{c}$) and anions (HN$_{a}$), and position of the first maximum of the
cation-water ($d_{c-O_{w}}$), and anion-water ($d_{a-O_{w}}$) in the radial
distribution function. In
parentheses, experimental data taken from the work of Marcus\cite{mar:cr88}.
	Properties were calculated at low concentrations and close to the solubility limit of each salt.}
\label{tab_res_cloruros}
  \begin{center}
    \begin{tabular}{ c c c c c c c c c}
\hline
\hline
	    Salt   & $m$  & CIP   & HN$_{c}$    & HN$_{a}$   & $d_{c-O_{w}}$ & $d_{S-O_{w}}$ & $d_{O_{s}-O_{w}}$ \\
	      &(mol/kg) &    &     &    &  \AA& \AA & \AA\\
\hline
	    Rb$_{2}$SO$_{4}$   & 0.5 & 0.35  & 6.2(5-8)     & 12.5(6.4-8.1)     & 2.75(2.79-2.90) & 3.76(3.67-3.89) &3.02(2.84-2.95)\\
	    & 1.5 & 0.70  & 6.0     & 11.7     & 2.75 & 3.77 &3.02\\
	    Cs$_{2}$SO$_{4}$  & 1  & 0.45  & 6.5(8-9)     & 12(6.4-8.1)     & 2.86(2.95-3.20) & 3.78(3.67-3.89) & 3.02(2.84-2.95)\\
	    & 5  & 1.25  & 5.8     & 10.2     & 2.85 & 3.76 & 3.02\\
\hline
\hline
    \end{tabular}
  \end{center}
\end{table*}

\subsection{Density of molten salts}

  Although the main purpose of using scaled charges is to improve the description of the aqueous 
  solution we shall now present results obtained for the molten salts (at room pressure and at 
  the experimental melting temperature). 

In table \ref{tab_dens_melt_solid} we have collected the results for the density of the
molten salts. As we pointed out in the Madrid-2019 force field, the use of scaled charges 
improves the description of the aqueous electrolyte solutions but at the cost of correctly describing 
the properties of the solid and of the melt. 
 In Table \ref{tab_dens_melt_solid} we can see
that the results obtained for the densities of the melt with scaled charges are about 
20 or even 25 $\%$ below the experimental molten densities. When we use a unit charge
(with the same parameters as for 0.85 charge) the results are similar to the
experimental ones. It is interesting to mention
that we have adjusted the cation-anion  and anion-anion interactions
using the results for the densities of the melt. These interactions have almost no 
effect in aqueous solution where by far the most important interactions are cation-oxygen and
anion-oxygen. However the parameters of the cation-anion interactions were determined having two 
properties in mind, namely the density of the melt and a reasonable number of CIP at the experimental value of the 
solubility limit. 
It should be mentioned that in the simulations  we found problems
for LiF and LiBr in getting a stable melt at the experimental conditions.  In the case of LiF we found spontaneous precipitation, 
and for LiBr spontaneous cavitation (due to the formation of chains of ions).
For these two salts our reported simulation were obtained at 200 K above the experimental melting temperature for LiF
and 3000 bar at the experimental melting temperature for LiBr (at 1 bar a 7-8 $\%$ lower density 
is expected taking into account
the experimental behavior of other molten salts)\cite{kcl-molten}.

\begin{table}[H]
\caption{Densities of the molten salts (at 1 bar and the experimental melting
temperature\cite{haynes2014}). Values of this work are given under columns labelled $q_{sc}$
	and $q$ which are obtained using scaled charges ($\pm$0.85 Z $e$)
	and integer charges ($\pm$1.0 Z $e$) for the
ions, respectively. For LiF the simulations results were obtained at  200 K
	above the experimental melting temperature. For LiBr the simulation results were obtained at 3000 bar. }
\label{tab_dens_melt_solid}
  \begin{center}
    \begin{tabular}{c c c c c c c c}
\hline
\hline
Salt  & \multicolumn{3}{c}{Melt density} \\
 & \multicolumn{3}{c}{kg/m$^{3}$} \\
\cline{2-4}
     & Expt & $q_{sc}$ & $q$  \\
\hline

	    LiF & 1810 & 1231  & 1570\\
	    LiBr &  2528 & 2220  &  2445\\
LiI & 3109& 2600  & 3087\\
NaF & 1948 & 1524  & 1949\\
NaBr & 2342 & 1954  & 2429\\
NaI & 2742 & 2166  & 2685\\
KF & 1910 & 1327  & 1708\\
KBr & 2127 & 1634  & 2090\\
KI & 2448 & 1825 &  2325\\
RbF & 2870 & 2260  & 2869\\
RbCl & 2248 & 1547  & 1986\\
RbBr & 2715 & 2150  & 2715\\
RbI & 2904 & 2279 &  2876\\
CsF & 3649 & 2864 &  3653\\
CsCl & 2790 & 1989 &  2568\\
CsBr & 3133 & 2357 &  3036\\
CsI & 3197 & 2479 &  3191\\
MgBr$_2$ & 2620 & 2305 & 2555\\
MgI$_2$ & 3050 & 2693 &  3004\\
CaBr$_2$ & 3111 & 2485 &  3017\\
CaI$_2$ & 3443 & 2766 &  3335\\

\hline
\hline
    \end{tabular}
  \end{center}
\end{table}

\subsection{Self diffusion coefficient of ions at infinite dilution}

\begin{center}
\begin{figure}[!hbt] \centering
    \centering
    \includegraphics*[clip,scale=0.3,angle=0.0]{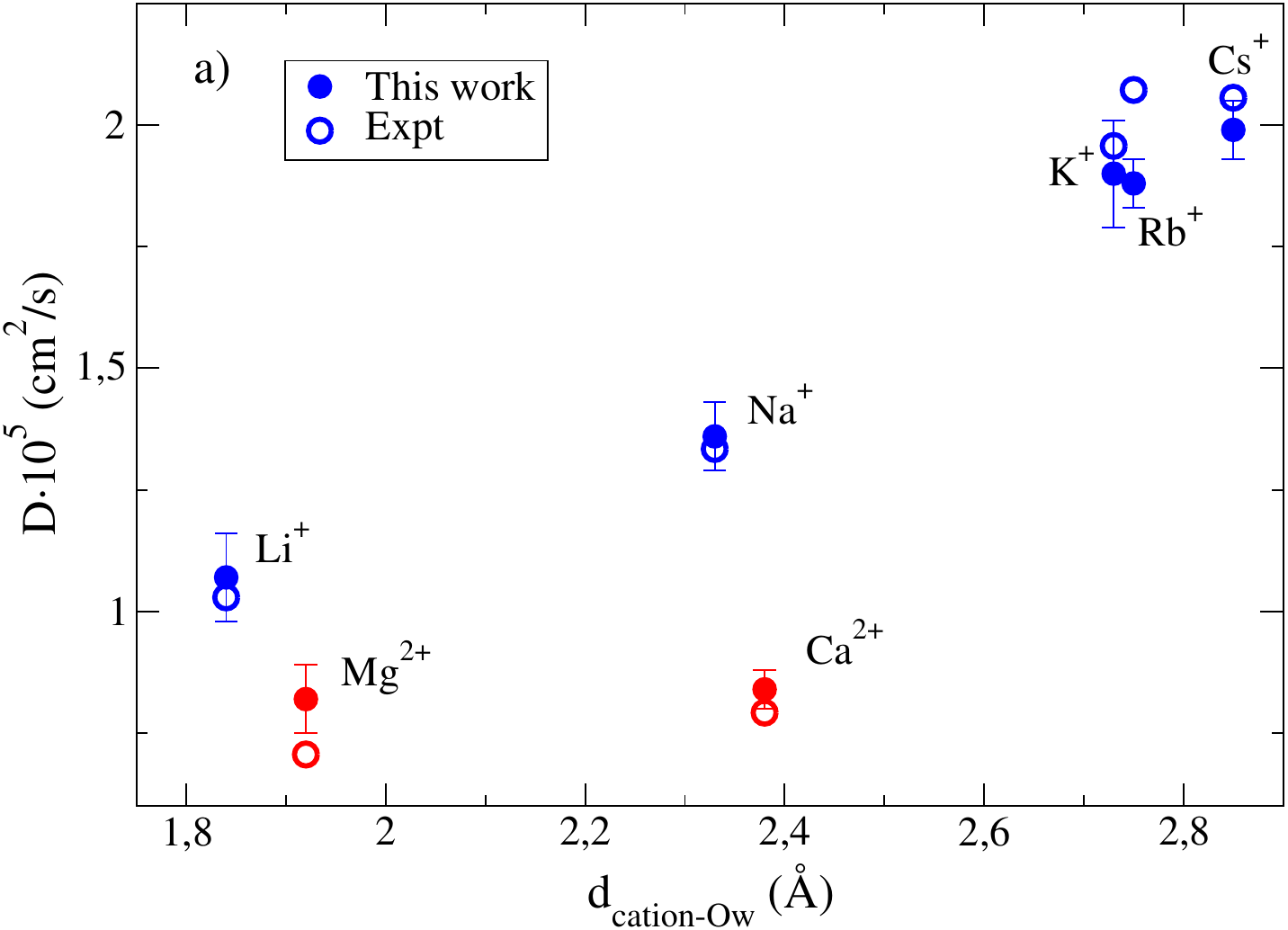}
    \includegraphics*[clip,scale=0.3,angle=0.0]{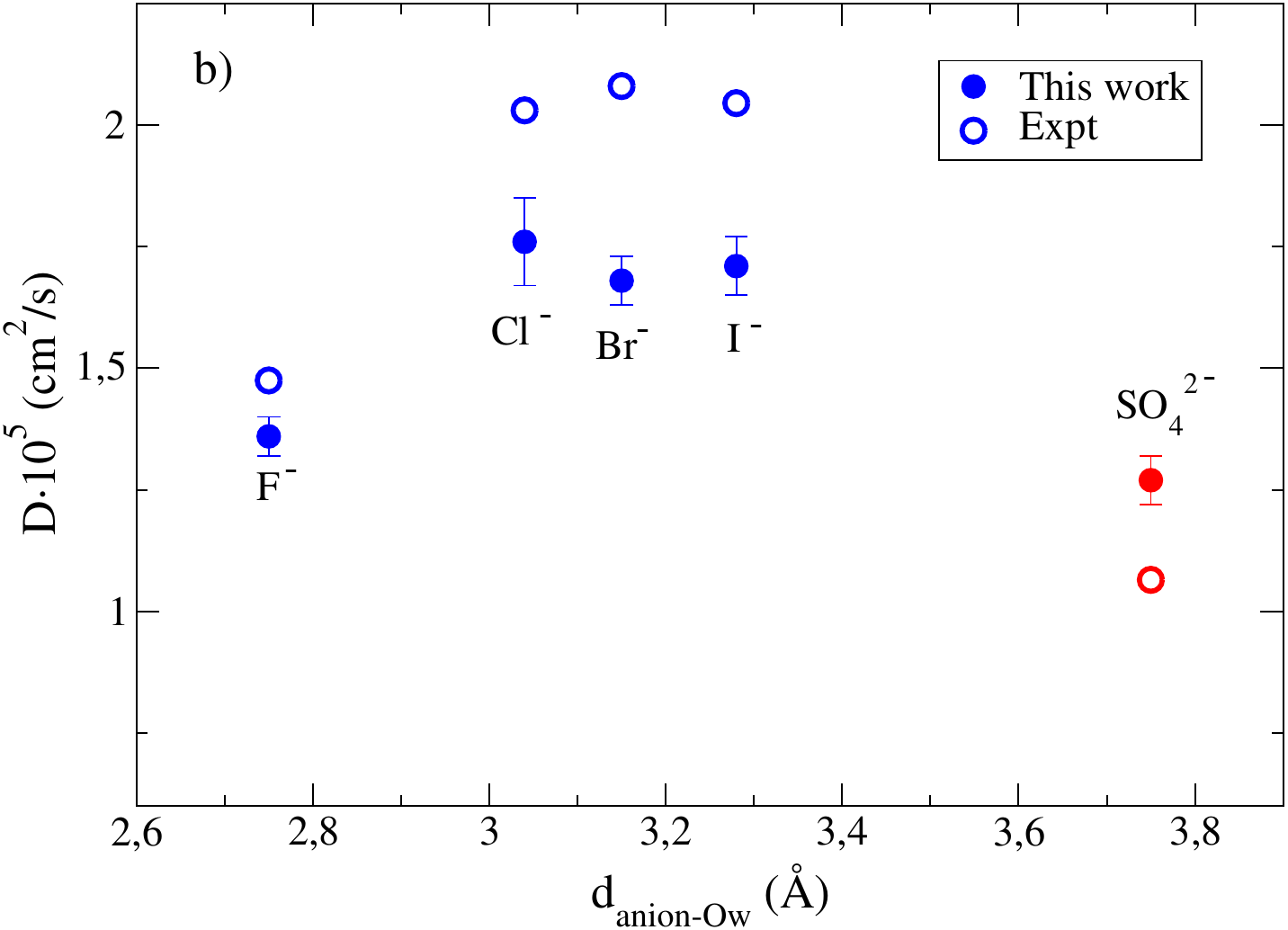}
	\caption{Diffusion coefficients of the Madrid-2019 ions at 1 bar and 298.15 K (full circles) compared with experimental values\cite{haynes2014} (empty circles)
	in function of the  position of the first maximum of the
cation-water ($d_{cation-O_{w}}$), and anion-water ($d_{anion-O_{w}}$) in the radial
distribution function. 
	a) Results for cations. In blue monovalent cations, in red, divalent cations. b) Results for anions. In blue monovalent anions, in red, sulfate anions.}
    \label{diffusion-coefficient}
\end{figure}
\end{center}

 Let us finish by presenting results for the diffusion coefficient of the ions at infinite dilution. 
We have performed molecular dynamics simulations of systems with 5550 water
molecules and 1 ion in order to study the self diffusion  coefficient of the ion
at infinite dilution. The fact that we are using only one ion (we have denoted that as
single ion) implies that the system is not electroneutral (although technically the use 
of Ewald sums implies that one has a neutralizing diffuse background of opposite charge). 
This is standard practice done to compute diffusion coefficients at infinite dilution. 
However, to check for the possible existence of artefacts we also simulated an electroneutral 
system having 5550 water molecules, 1 Na$^+$ and 1 Cl$^-$ corresponding to a molality of 
0.01 m  and we obtained quite similar results to those obtained with single
ion simulations. 
Thus, we have used the single ion method in the rest of the ions. All the results collected in 
Table \ref{tabla-iones} and presented in Figure \ref{diffusion-coefficient}
are the result of the average of 5 independent runs of 40 ns with 5550 water 
molecules and 1 ion. So that, each point obtained is the result of 200 ns of 
simulation, 2.8 $\mu$s in total. For the Madrid-2019 model we have also included in Table 
\ref{tabla-iones} the results obtained by Dopke et $al.$\cite{dopke}, obtaining good agreement except 
for Cl$^-$ where we obtained an slightly larger value (although still within the combined uncertainty).
Notice that the results of Dopke et $al.$ were obtained using a smaller system having 523 molecules 
of water.

\begin{table}[!hbt] \centering
	\caption{Diffusion coefficients of the Madrid-2019 ions (in
	cm$^{2}$/s) at 1 bar and 298.15 K. We 
present the results obtained in this work, the results of Dopke et $al.$
\cite{dopke}.  
The experimental values are from \cite{haynes2014}. 
The results  are obtained from the
average of five independent simulations and we have applied the hydrodynamic corrections of Yeh and Hummer\cite{Yehhummer}.}
\label{tabla-iones}
  \begin{center}
    \begin{tabular}{c c c c c c c c c c c c c c c}
\hline
\hline
Ion & & $D_{exp}$$\cdot$10$^{5}$       &  $D_{this\;work}$$\cdot$10$^{5}$   & $D_{this\;work}$$\cdot$10$^{5}$ & $D_{Dopke}$$\cdot$10$^{5}$
                                    \\
                                         & &  &   Single Ion  &Electroneutral& \\
\hline 
Na$^{+}$ & & 1.334 &   1.36(07) &  1.39(11) & 1.28(15)    \\
Cl$^{-}$ & & 2.03 &   1.76(09) &  1.75(08) & 1.60(08) \\
K$^{+}$ & & 1.957 &  1.90(11) &  - &  1.93(06)   \\
Li$^{+}$ & & 1.029 &   1.07(09) &  -  &  1.08(02)   \\
Mg$^{2+}$ & & 0.706&   0.82(07) &  -  &  0.87(13)   \\
Ca$^{2+}$ & & 0.792 &   0.84(04) &  -  &  0.89(08)   \\
SO$_4$$^{2-}$ & & 1.065 &   1.27(05) &  -  &  -   \\
F$^{-}$ & & 1.475 &   1.36(4) &  - & - \\
Br$^{-}$ & & 2.080 &   1.68(5) &  - & - \\
I$^{-}$ & & 2.045 &   1.71(6) &  - & - \\
Rb$^{+}$ & & 2.072 &   1.88(5) &  - & - \\
Cs$^{+}$ & & 2.056 &   1.99(6) &  - & - \\
\hline
\hline

    \end{tabular}
  \end{center}
\end{table}

Figure \ref{diffusion-coefficient}a) shows that the results
for most of the cations 
(with the exception of Rb$^{+}$) are in excellent agreement with experimental
results. It can also be seen that for monovalent cations the diffusion coefficient
increases as the water-cation distance increases. For divalent cations,
on the other hand, there is no major change in the diffusion coefficient
with increasing water-cation distance. 
In the case of the anions (in particular for halogens) the results
obtained are below the experimental ones with the exception of the sulfate in which 
the opposite is true as we can see in Figure \ref{diffusion-coefficient}b).
In this case the diffusion coefficient increases from fluorine
anion to chlorine anion and remains constant thereafter.
Notice that we have applied the
hydrodynamic corrections of Yeh and Hummer\cite{Yehhummer}.
using the viscosiy of the TIP4P/2005 water model. We have
considered that the viscosity of the systema it is not
affected by the dilution of 1 single ion (i.e. 
at infinite dilution.

\subsection{Diffusion coefficient of water for several salts: Stokes-Einstein relation for aqueous electrolytes }

\begin{center}
\begin{figure}[!hbt] \centering
    \centering
    \includegraphics*[clip,scale=0.3,angle=0.0]{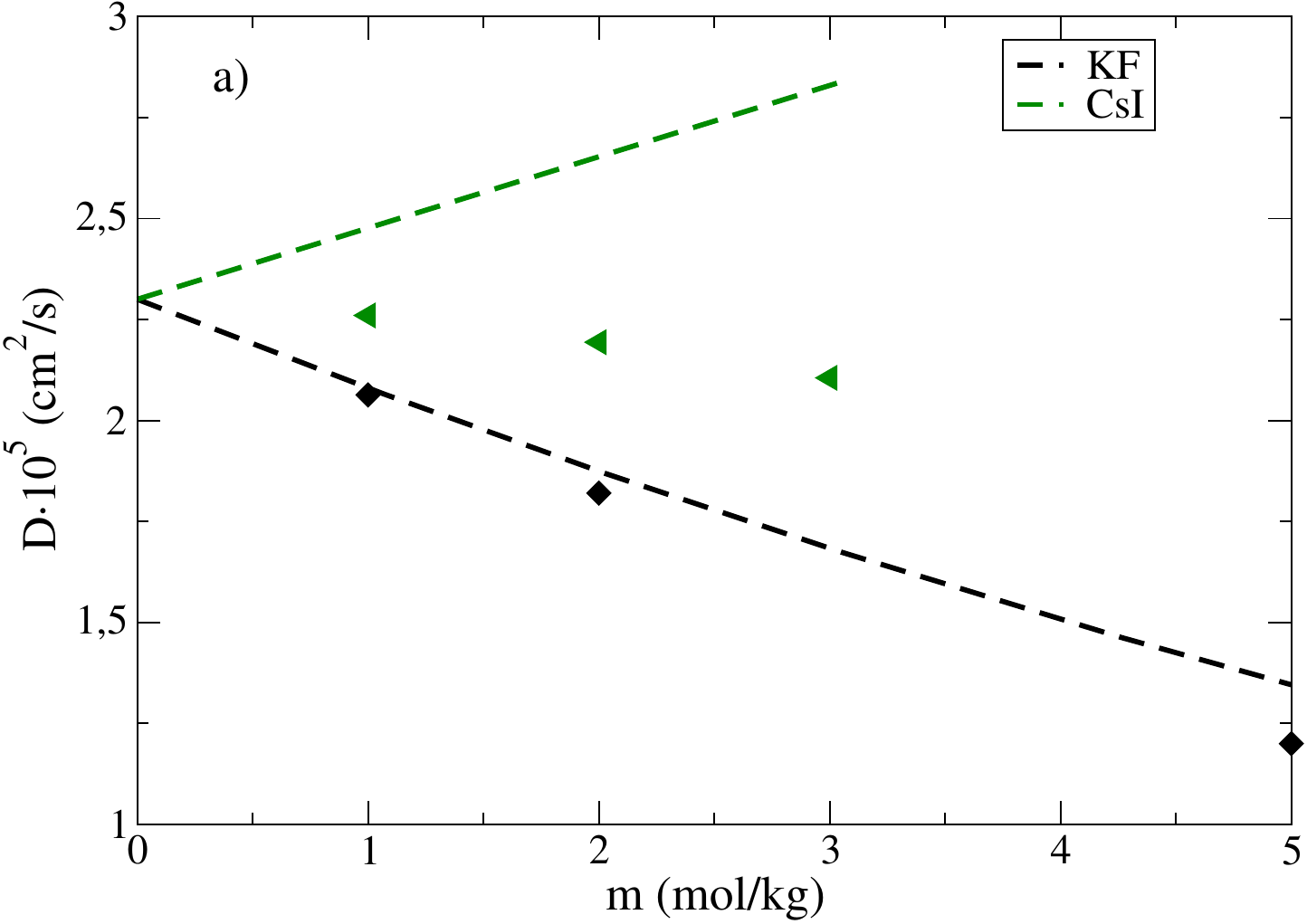}
    \includegraphics*[clip,scale=0.3,angle=0.0]{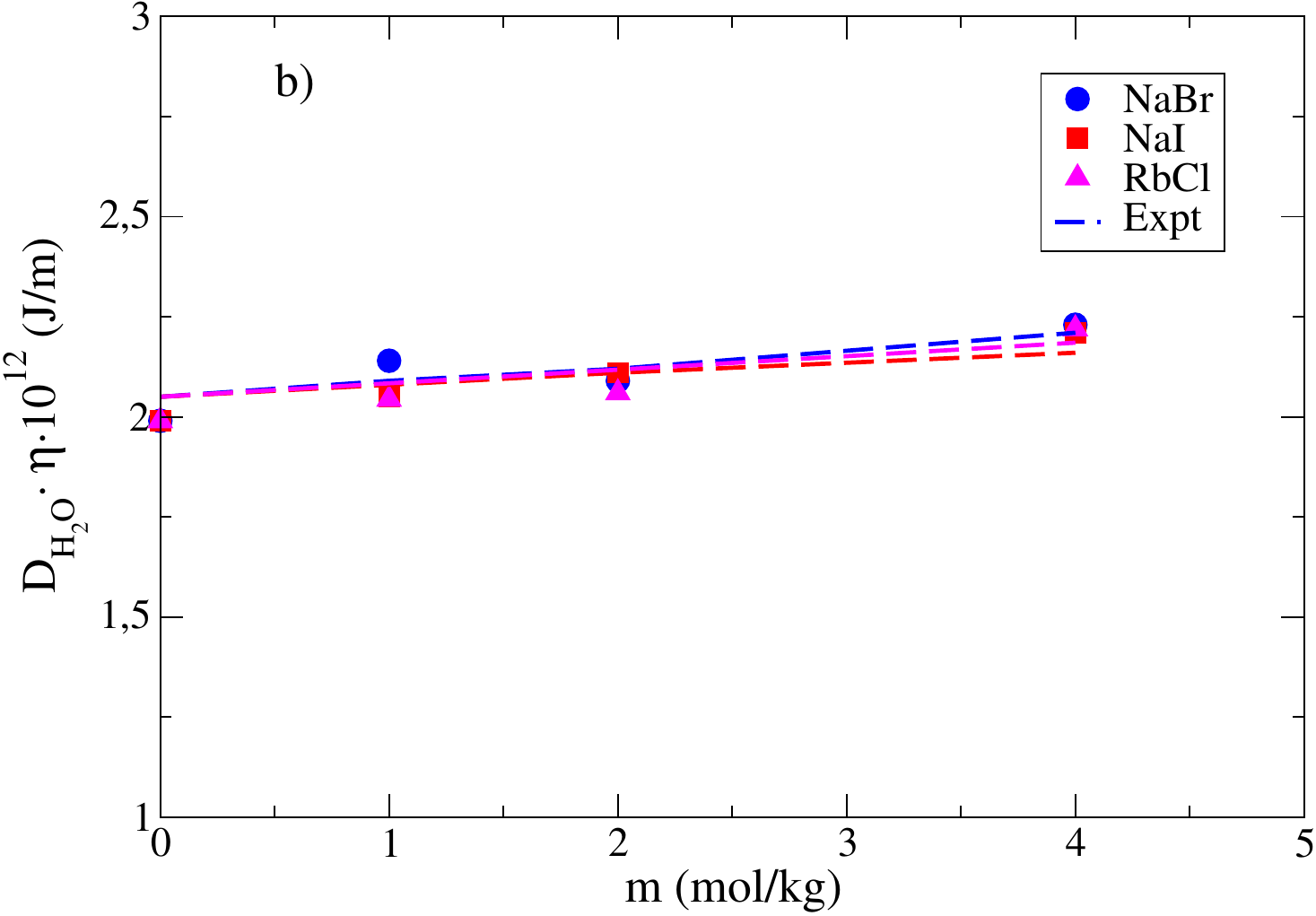}
	\caption{a) Diffusion coefficients of water for several salts
	of the Madrid-2019-Extended at 1 bar and 298.15 K. Green triangles are
	the results for CsI and black squares for KF.
	b)Product of the water diffusion coefficient and viscosity
        in function of molality for different salts of the Madrid-2019-Extended model: Blue circles
        are the results for NaBr, red squares for NaI and pink triangles for RbCl. Experimental
        results are plotted in dashed lines in the same colors as their respective salts.
        Experimental diffusion coefficients obtained from Ref\cite{mul:jpc96}.}
    \label{diffusion-water}
\end{figure}
\end{center}

In this work we have also considered it relevant to analyse the water diffusion coefficients
for various aqueous solutions of electrolytes at different concentrations.
The salts chosen have been some of those for which we have calculated the viscosity.
This is due to the fact that when applying the Yeh and Hummer corrections\cite{Yehhummer}, the viscosity used was that of the model.
As can be seen in Figure \ref{diffusion-water}a, the force field, in the case of KF, is able to reproduce the experimental results 
accurately (it was the salt that best reproduced the viscosities).
Nevertheless, in the case of CsI we have the same problem that other authors pointed out \cite{kim12}. 
Diffusion coefficient of water in CsI aqueous solutions increases with the concentration. This behaviour
is anomalous and molecular dynamics simulations are not able (even using scaled charges).
For CsI the model  is able to capture at least a very small impact of the salt in the diffussion coefficient of water.  
It is true that Ding et \textit{al.}\cite{Ding3310} showed that with ab initio calculations
it is possible to reproduce this trend.
We have also calculated the diffusion coefficients 
of water in the presence of other three salts in order to evaluate if the product of
the viscosity and the diffusion coefficient of water remains constant (i.e. 
if the Stokes-Einstein relation\cite{dill2003molecular} is satisfied).
In Figure \ref{diffusion-water}b we have plotted the values of the product 
of the water diffusion coefficient and viscosity obtained in this work
versus salt concentration. It can be seen how the all the results follow the same trend 
and  all salts keep the $D\cdot\eta$ product almost constant with a slightly increase with
the concentration. The experimental results
(dashed lines) show the same trend. The agreement between experimental and simulation
results for the product $D\cdot\eta$ is excellent.
Of course, just because the model reproduces the experimental results of the product
$D\cdot\eta$ does not mean that it reproduces the experimental values of viscosity and diffusion coefficients.
What happens is that when the force field overestimates the experimental viscosity,  
to keep constant the product (and reproduce experimental results), the model must underestimate 
the diffusion coefficients of the water.

\newpage
\section{Concluding remarks}

In this work we have developed the Madrid-2019-Extended force field for electrolytes 
in water (as described by the TIP4P/2005 model). This force field extends the 
Madrid-2019 force field to the cations Rb$^{+}$ and Cs$^{+}$ and to the anions F$^{-}$, Br$^{-}$ and I$^{-}$. 
Thus, the Madrid-2019-Extended force field includes the ions considered in the 
celebrated Joung-Cheatham force field, and some additional ions such as Mg$^{2+}$ and Ca$^{2+}$. 
We have presented results for the densities of a number of salts, and of the viscosities
for some selected salts. Hydration numbers, radial distribution functions, and contact ion 
pairs were also calculated as well as diffusion coefficients of the ions at infinite dilution. 
The main conclusions of this work are as follows:

\begin{itemize}
\item{The use of scaled charges allows us to describe accurately the densities of 
a large number of salts as was already the case with the Madrid-2019 force field.}
\item{As we have pointed out in the Madrid-2019 force field, the use of scaled charges 
improves the description of aqueous solutions but the cost is the 
description of solid phases: densities
for molten salts are underestimated by about a 20$\%$}
\item{Following the philosophy of the Madrid-2019 force field the charge 0.85 Z $e$ describes
quite well the viscosities for concentrations up to 2 m. However, at higher molalities
		there are  large deviations, especially for divalent salts.} 
\item{Self diffusion coefficients at infinite dilution are described accurately for most
of the cations but not for the anions, whose results are slightly worse.}
\item{No spontaneous precipitation was found when performing simulations at the experimental
	value of the solubility limit. }
\item{ The number of CIP was in general below 0.5 for salts with solubility smaller than 10 m. 
	For salts with huge solubility this rule can not be applied. 
		For these salts the number of CIP was smaller by about a factor of between 
		0.6-0.8 of that found from the random mixing rule, thus illustrating that 
		water is a good solvent for the ions. }
\end{itemize}

 To summarize, the combination of a good model of water and scaled charges 
 yields a reasonable description of electrolyte solutions improving unit charges
 force fields results.
 The Madrid-2019-Extended should be regarded as a computationally cheap way 
 of introducing some degree of polarization. The model provides reasonable results 
 but for certain properties the agreement with experiment is not quantitative thus 
 showing the limits of this approach. In the particular case of transport properties 
 it is clear that there is room for improvement. One could argue that polarizable models 
 should improve the description. Time is needed to provide evidence of that. 
 Also, the impact of nuclear quantum effects
 in transport properties of electrolytes has not been considered in detail in the literature. 
 For the time being we hope that the Madrid-2019-Extended, with care, could be useful 
 to provide some predictions regarding interesting physical problems. 
  In the future it will be useful to study the performance of the Madrid-2019 force field in 
  a number of problems such as solubilities, freezing point depression, nucleation of ice in 
  salty solutions and electrical conductivity. A face to face comparison with force fields 
  that use integer charges will bring more evidence of the benefits and drawbacks of using 
  scaled charges. 

\section{Supplementary}
In the supplementary material we have collected the numerical results for densities
and viscosities obtained in this work for several salt solutions and the 
complete set of parameters for the force field Madrid-2019 and 
its extended version. See also topol.top
file of GROMACS with the force field Madrid-2019 and its extended version
attached with this paper.

\section*{Acknowledgments}

This work was funded by Grants PID2019-105898GB-C21 and PID2019-105898GA-C22 
of the MICINN and GR-910570 (UCM). 
M.M.C. acknowledges CAM and UPM for financial support of this work through the CavItieS project
No. APOYO-JOVENES-01HQ1S-129-B5E4MM from 
``Accion financiada por la Comunidad de Madrid en el marco del Convenio Plurianual con la
Universidad Politecnica de Madrid en la linea de actuacion estimulo a la investigacion de jovenes doctores''.
S.B. thanks Ministerio de Educacion y Cultura for a pre-doctoral FPU Grant No. 
FPU19/00880. 
The authors gratefully acknowledge the Universidad Politecnica de Madrid (www.upm.es)
for providing computing resources on Magerit Supercomputer.

\section*{Conflict of interest}
The authors declare that there are no conflicts of interest.

\section*{Data Availability}
The data that support the findings of this work are available
within the article and its supplementary material.

 %\bibliographystyle{apsrev}
%\bibliography{paper-doble-columna.bbl}

\end{document}